\documentclass[superscriptaddress,
floatfix,
reprint,
nofootinbib,
amsmath,amssymb,
aps,
prd,
]{revtex4-2}

\usepackage{graphicx}

\usepackage{bm}
\usepackage{hyperref}
\usepackage{cleveref}

\usepackage{dcolumn}
\usepackage{booktabs} 
\usepackage{tabularx}
\usepackage{multirow}
\renewcommand{\arraystretch}{1.5}
\usepackage{threeparttable}

\usepackage{chemformula}

\usepackage{gensymb} 
\usepackage{mycommands}
\usepackage{dblfloatfix}
\usepackage{placeins}

\usepackage{siunitx}
\DeclareSIUnit\torr{torr}
\DeclareSIUnit\ppm{ppm}
\DeclareSIUnit\ppb{ppb}
\DeclareSIUnit\bar{bar}
\DeclareSIUnit\gauss{G}
\DeclareSIUnit\neV{\nano\electronvolt}
\DeclareSIUnit\peV{\pico\electronvolt}
\DeclareSIUnit\kms{km/s}
\usepackage[normalem]{ulem}
\usepackage{changes}

\begin{document}

\preprint{APS/123-QED}

\title{Search for Axionlike Dark Matter Using Liquid-State Nuclear Magnetic Resonance}

\author{Julian Walter}
\thanks{These authors contributed equally to this work.}
\affiliation{Johannes Gutenberg-Universit{\"a}t Mainz, 55128 Mainz, Germany}
\affiliation{Helmholtz Institute Mainz, 55099 Mainz, Germany}
\affiliation{GSI Helmholtzzentrum für Schwerionenforschung GmbH, 64291 Darmstadt, Germany}

\author{Olympia Maliaka}
\thanks{These authors contributed equally to this work.}
\affiliation{Johannes Gutenberg-Universit{\"a}t Mainz, 55128 Mainz, Germany}
\affiliation{Helmholtz Institute Mainz, 55099 Mainz, Germany}
\affiliation{GSI Helmholtzzentrum für Schwerionenforschung GmbH, 64291 Darmstadt, Germany}

\author{Yuzhe Zhang}
\thanks{These authors contributed equally to this work.}
\affiliation{Johannes Gutenberg-Universit{\"a}t Mainz, 55128 Mainz, Germany}
\affiliation{Helmholtz Institute Mainz, 55099 Mainz, Germany}
\affiliation{GSI Helmholtzzentrum für Schwerionenforschung GmbH, 64291 Darmstadt, Germany}

\author{John W. Blanchard}
\affiliation{Quantum Technology Center, University of Maryland, College Park, MD 20742, United States of America}
\affiliation{Helmholtz Institute Mainz, 55099 Mainz, Germany}
\affiliation{GSI Helmholtzzentrum für Schwerionenforschung GmbH, 64291 Darmstadt, Germany}

\author{Gary Centers}
\affiliation{Johannes Gutenberg-Universit{\"a}t Mainz, 55128 Mainz, Germany}
\affiliation{Helmholtz Institute Mainz, 55099 Mainz, Germany}
\affiliation{GSI Helmholtzzentrum für Schwerionenforschung GmbH, 64291 Darmstadt, Germany}

\author{Arian Dogan}
\affiliation{Johannes Gutenberg-Universit{\"a}t Mainz, 55128 Mainz, Germany}
\affiliation{Helmholtz Institute Mainz, 55099 Mainz, Germany}
\affiliation{GSI Helmholtzzentrum für Schwerionenforschung GmbH, 64291 Darmstadt, Germany}

\author{Martin Engler}
\affiliation{Johannes Gutenberg-Universit{\"a}t Mainz, 55128 Mainz, Germany}
\affiliation{Helmholtz Institute Mainz, 55099 Mainz, Germany}
\affiliation{GSI Helmholtzzentrum für Schwerionenforschung GmbH, 64291 Darmstadt, Germany}

\author{Nataniel L. Figueroa}
\affiliation{Johannes Gutenberg-Universit{\"a}t Mainz, 55128 Mainz, Germany}
\affiliation{Helmholtz Institute Mainz, 55099 Mainz, Germany}
\affiliation{GSI Helmholtzzentrum für Schwerionenforschung GmbH, 64291 Darmstadt, Germany}

\author{Younggeun Kim}
\affiliation{Johannes Gutenberg-Universit{\"a}t Mainz, 55128 Mainz, Germany}
\affiliation{Helmholtz Institute Mainz, 55099 Mainz, Germany}
\affiliation{GSI Helmholtzzentrum für Schwerionenforschung GmbH, 64291 Darmstadt, Germany}

\author{Derek F. Jackson Kimball}
\affiliation{Department of Physics, California State University--East Bay, Hayward, CA 94542-3084, United States of America}

\author{Matthew Lawson}
\affiliation{Physics Department, Stockholm University, Albanova University Centre, SE-10691 Stockholm, Sweden}

\author{Declan W. Smith}
\affiliation{Department of Physics, Boston University, Boston, MA 02215, United States of America}

\author{Alexander O. Sushkov}
\affiliation{Department of Physics, Boston University, Boston, MA 02215, United States of America}
\affiliation{Department of Electrical and Computer Engineering, Boston University, Boston, MA 02215, United States of America}
\affiliation{Photonics Center, Boston University, Boston, MA 02215, United States of America}
\affiliation{Department of Physics \& Astronomy, The Johns Hopkins University, Baltimore, MD 21218, United States of America}

\author{Dmitry Budker}
\affiliation{Johannes Gutenberg-Universit{\"a}t Mainz, 55128 Mainz, Germany}
\affiliation{Helmholtz Institute Mainz, 55099 Mainz, Germany}
\affiliation{GSI Helmholtzzentrum für Schwerionenforschung GmbH, 64291 Darmstadt, Germany}
\affiliation{Department of Physics, University of California, Berkeley, CA 94720-7300, United States of America}

\author{Hendrik Bekker}
\email{hbekker@uni-mainz.de}
\affiliation{Johannes Gutenberg-Universit{\"a}t Mainz, 55128 Mainz, Germany}
\affiliation{Helmholtz Institute Mainz, 55099 Mainz, Germany}
\affiliation{GSI Helmholtzzentrum für Schwerionenforschung GmbH, 64291 Darmstadt, Germany}

\author{Arne Wickenbrock}
\affiliation{Johannes Gutenberg-Universit{\"a}t Mainz, 55128 Mainz, Germany}
\affiliation{Helmholtz Institute Mainz, 55099 Mainz, Germany}
\affiliation{GSI Helmholtzzentrum für Schwerionenforschung GmbH, 64291 Darmstadt, Germany}


\begin{abstract}
We search for dark matter in the form of axionlike particles (ALPs) in the mass range $\SI{5.576741}{\neV/c^2}-\SI{5.577733}{\neV/c^2}$ by probing their possible coupling to fermion spins through the ALP field gradient. This is achieved by performing proton nuclear magnetic resonance spectroscopy on a sample of methanol as a technical demonstration of the Cosmic Axion Spin Precession Experiment Gradient (CASPEr-Gradient) Low-Field apparatus. Searching for spin-coupled ALP dark matter in this mass range with associated Compton frequencies in a \SI{240}{\Hz} window centered at \SI{1.348570}{\MHz} resulted in a sensitivity to the ALP-proton coupling constant of $g_{\mathrm{ap}} \approx 3 \times 10^{-2}\,\mathrm{GeV}^{-1}$. This narrow-bandwidth search serves as a proof-of-principle and a commissioning measurement, validating our methodology and demonstrating the experiment's capabilities. CASPEr-Gradient Low-Field will probe the mass range from \SI{4.1}{\peV/c^2} to \SI{17}{\neV/c^2} with hyperpolarized samples to boost the sensitivity beyond the astronomical limits.
\end{abstract}

\maketitle

\section{Introduction}

The hypothesis of dark matter (DM) is well supported by astronomical and cosmological observations. 
Early examples include studies of galaxy clusters by Zwicky \textit{et al.} in the 1930s\,\cite{zwicky1933redshift,zwicky1979masses} and of galactic rotation curves by Rubin \textit{et al.} in the 1970s\,\cite{rubin1970rotation}.
Since then, observations supporting the DM hypothesis\,\cite{ade2016planck, bartelmann2010gravitational,Clowe_2006,Harvey_2015,jedamzik2009big} have been accumulating.
Among various well-motivated DM candidates, axions and axionlike particles (ALPs)\,\cite{berezhiani1992primordial, annurevAxionSearches, chadhaday2022axiondarkmatternow, Kimball2023_search} are the focus of this work. 
These ultralight pseudoscalar bosons with possible masses between roughly $10^{-22}$ and $\SI{10}{\eV/c^2}$\,\cite{GrahamStochasticAxion} are introduced by theoretical models of particle physics such as string theory and grand unified theories (GUT)\,\cite{Preskill1983_invisible_axion, Abbott1982_Cosmological_Axion, Dine1983_not_so_harmless_axion,svrcek2006axions, berezhiani1991cosmology}.
Axions are particularly of interest as they emerged as a solution to the strong-$CP$ problem of quantum chromodynamics\,\cite{PecceiQuinn1977_CPConservation_PRL, PecceiQuinn1977_ConstraintsCP_PRD}. 
For convenience, we refer to both axions and axionlike particles as ALPs unless there is a need to distinguish them.

Considering the local DM density $\rho_{\text{DM}} \approx \SI{0.3}{\GeV/\cm^{3}}$ \cite{BERGSTROM1998137, Jungman1992SupersymmetricDM, doi:10.1146/annurev.astro.39.1.137} and the ALP mass, if ALPs make up a significant fraction of DM, their number density is large enough that their collective behavior can be approximately described as a classical wave: 
\begin{equation}\label{eq:ALPfield}
    a(\mathbf{r},t) = a_0 \cos{(2\pi\nu_a t - \mathbf{k}\cdot\mathbf{r}+\phi)}\,,
\end{equation}
where $a_{0}$ is the field amplitude, $\mathbf{r}$ is the displacement vector; $\mathbf{k}=m_a\bm{v}_a/\hbar$ is the wave vector with ALP velocity $\bm{v}_a$, ALP mass $m_a$ and the reduced Planck constant $\hbar$; $\phi$ is a random phase\,\cite{Foster2018HaloAxion,Lisanti2021Stoch,Cheong2025QuantumDM}.
The field oscillates at a frequency close to the Compton frequency $\nu_{a}= m_{a}c^2/(2\pi\hbar)$, where $c$ is the speed of light in vacuum.
Following the standard halo model (SHM) and assuming that DM is virialized in the Milky Way\,\cite{Primack1988DetectionOC}, the value of $a_0$ can be estimated from the DM energy density $\rho_{\text{DM}}\approx{c m_a^2 a_0^2}/{(2\hbar^3)}$\,\cite{kimball2018overview}. 
The velocity of DM bound within the galactic halo follows a Maxwell-Boltzmann distribution with an average speed of $v_0\approx$\,\SI{220}{km/s}, an isotropic velocity dispersion of $v_\sigma \approx \SI{156}{\km/\s}$\,\cite{Primack1988DetectionOC}, and a cut-off at the escape velocity of $v_\mathrm{esc}\approx\SI{544}{km/s}$\,\cite{Evans2019_SHMRefined}.
Taking the velocity distribution into account, a background field of axionlike DM is more accurately expressed as a sum over $N$ independent oscillators, each of the form of Eq.\,\eqref{eq:ALPfield}, where the velocity of the $n$-th ALP is sampled from the distribution of velocities in the halo.
The distribution of ALP velocities results in a spread of ALP frequencies. 
With a mean velocity dominated by Earth's motion, a terrestrial detector observes the distribution of ALP frequencies with a mode of $\approx \nu_a[1+v_0^2/(2c^2)]$\,\cite{kimball2018overview}. 
Individual ALP waves of frequency $\nu_n$ and phase $\phi_n$ interfere with each other and the resulting amplitude follows a Rayleigh distribution\,\cite{Foster2018HaloAxion, Lisanti2021Stoch, Gramolin2022Feb_SpectralSignatures}.
Therefore, the ALP field amplitude and phase vary randomly on a timescale of the characteristic coherence time $\tau_a \approx (\nu_a v_a^2 / c^2)^{-1}$\,\cite{Schive2014,  Centers2021Stochastic, Gramolin2022Feb_SpectralSignatures}.
As it affects the computed or measured experimental sensitivity, this stochastic behavior of the ALP signal is taken into account during analysis. 

There are three possible non-gravitational couplings between ALPs and Standard Model (SM) particles: (a) coupling to photons, (b) coupling to gluons and (c) coupling to fermion spins, also referred to as the gradient coupling\,\cite{Kimball2023_search}. Various experiments\,\cite{ouellet2019first,du2018search,ALPHA2023,aja2022canfranc,CAPP2024,brouwer2022projected,alesini2023future,zhong2018results,garcia2024searchaxiondarkmatter,quiskamp2024near,rettaroli2024search,gramolin2021search,gavilan2024searching,bloch2105nasduck,wu2019search,abel2023search,karanth2023first,lee2023laboratory,capozzi2023new} search for these couplings.

The Cosmic Axion Spin Precession Experiments (CASPEr), located in Boston, USA and Mainz, Germany, focus on the latter two couplings to search for ALPs\,\cite{CASPErProposal2014, kimball2018overview, wang2018application, wu2019search, aybas2021_SolidStateNMR, CASPEr_ZULF_2019}. 
In this paper, we report on measurements done with the CASPEr-Gradient Low-Field (LF) apparatus, a magnetometry-based experiment that investigates the interaction between the gradient of the ALP field and atomic spins.
While experiments looking for all three couplings are sensitive to stochastic fluctuations in the amplitude of the ALP field, a key distinction of experiments investigating the gradient coupling is their additional sensitivity to fluctuations in ALP velocity. 
The Hamiltonian of the gradient interaction can be written as:
\begin{equation}
   H_{\mathrm{aNN}}=g_{\mathrm{aNN}}\bm{\nabla} a\cdot \mathbf{I}\,, 
   \label{eq:axionham}
\end{equation}
where $g_{\mathrm{aNN}}$ is the coupling constant and $\mathbf{I}$ is the nuclear spin. 
In this paper, we focus on a measurement using a spin-1/2 proton sample. 
Thus, we are effectively probing the ALP gradient-proton coupling constant, $g_{\mathrm{ap}}$. 
The magnitude of the gradient can be estimated as $|\bm{\nabla}a|\,\approx a_0 m_a v_a/\hbar$. 
The Hamiltonian in Eq.\,\eqref{eq:axionham} is similar to that of the Zeeman interaction: $H=-\hbar \gamma~\mathbf{B}\cdot \mathbf{I}$, where $\gamma$ denotes the proton gyromagnetic ratio and $\mathbf{B}$ is the applied magnetic field. 
In analogy to this, one can define an oscillating ``pseudo-magnetic field'' generated by the gradient of the ALP field
\begin{equation}
    \mathbf{B_a}(t)=g_{\mathrm{ap}}\sqrt{2\hbar c\rho_{\text{DM}}}\,\frac{ \bm{v_a}\,\sin(2\pi\nu_a t)}{\gamma}\,.
    \label{eq:pseudoB}
\end{equation}
To detect the presence of such a pseudo-magnetic field, we use nuclear magnetic resonance (NMR) techniques. 
For an ensemble of nuclear spins in a bias magnetic field $\mathbf{B}_0$, the Larmor frequency of the spins $\nu_L$ is proportional to the strength of the bias field: $\nu_L=\gamma B_0/(2 \pi)$.
If the Larmor frequency matches the Compton frequency of the ALP background, the component of the ALP-field gradient perpendicular to $\mathbf{B_0}$, $\bm{\nabla} a_\perp$, serves as a continuous oscillating driving field, analogous to continuous-wave (cw) NMR.
As with a conventional cw-NMR experiment, this results in a transverse magnetization oscillating at the Larmor frequency which could be detected using, for example, lock-in amplification.
Note that CASPEr-Gradient, and any experiment probing the ALP-nucleon coupling via NMR, is by nature also sensitive to real magnetic fields generated by the presence of DM near the sample, such as dark photons\,\cite{CASPEr_ZULF_2019,beadle2025dark}.

The CASPEr-Gradient-LF has a tunable magnet producing $\mathbf{B_0}$ from zero to \SI{0.1}{\tesla}. 
For ALP searches in a higher mass range the CASPEr-Gradient High-Field (HF) apparatus is employed with a tunable zero-to-\SI{14.1}{\tesla} magnet and is in the calibration stage at the time of this work.
The sensitivity of the setup is nearly frequency-independent in the range of \SI{1}{\kHz} to \SI{4.3}{\MHz}.
In this work, we report on a search for the ALP-proton gradient coupling over a \SI{240}{\Hz} interval centered around \SI{1.348570}{\MHz} ($|\mathbf{B_0}|\approx\SI{31.67331}{\milli\tesla}$). This corresponds to ALP masses in a range of $\SI{0.992}{\peV/c^2}$ around $m_a\approx\SI{5.577237}{\neV/c^2}$.
We use a thermally-polarized sample and focus on a narrow bandwidth to validate our methodology\,\cite{CASPErProposal2014,garcon2017cosmic} and characterize the CASPEr-Gradient-LF apparatus as a tool to search for ALP DM. 

\section{Apparatus and sample}

The CASPEr-Gradient-LF apparatus includes a flow-through liquid-helium cryostat containing a superconducting solenoid that can generate a bias field of up to $B_0 =\SI{0.1}{\tesla}$. The DM mass range that can be searched using the CASPEr-Gradient apparatuses depends on the gyromagnetic ratio of the sample in use and on $B_0$. 
For example, see Table\,\ref{tab:parameters} for methanol and xenon (Xe) samples.
As evident from Table\,\ref{tab:parameters}, the degree of polarization of the sample presents the largest limiting factor to CASPEr-Gradient-LF sensitivity. 
For that reason, in the future we will employ a hyperpolarized liquid Xe sample, which can boost sensitivity by several orders of magnitude alone.

Eight superconducting shim coils surrounding the solenoid are used to reduce the field inhomogeneity to less than \SI{2}{\ppm} over a spherical volume with a radius of \SI{4}{\mm}. However, in the measurements, an inhomogeneity of approximately \SI{10}{\ppm} was obtained after tuning the shim fields using an algorithm based on Bayesian optimization\,\cite{ShimmingPaper}. This is larger than the \SI{2}{\ppm} inhomogeneity mentioned because we used a larger sample volume with a height of \SI{24}{\mm} and a radius of \SI{4}{\mm} and introduced an additional superconducting pickup coil. 
In general, higher number of spins results in better sensitivity. Therefore we chose liquid methanol (\ch{CH3OH}) as the sample due to its high proton density (\SI{99}{\mmol/\cm^3}), resulting in a total of $\SI{7.2e22}{}$ protons, and low freezing point (\SI{175.6}{\kelvin}) for it to remain liquid in the cold sample holder.
A variable temperature insert (VTI) is used to position the sample in the homogeneous region of the bias field. 
The sample was kept above the freezing point (verified via repeated pulsed-NMR measurements) by vacuum insulation from the liquid helium and by blowing warm nitrogen gas through a spiral surrounding the sample holder. 
A system capable of more advanced temperature monitoring and control is currently being implemented for future works.

\begin{figure}[ht]
    \centering
    \includegraphics[width=1.0\linewidth]{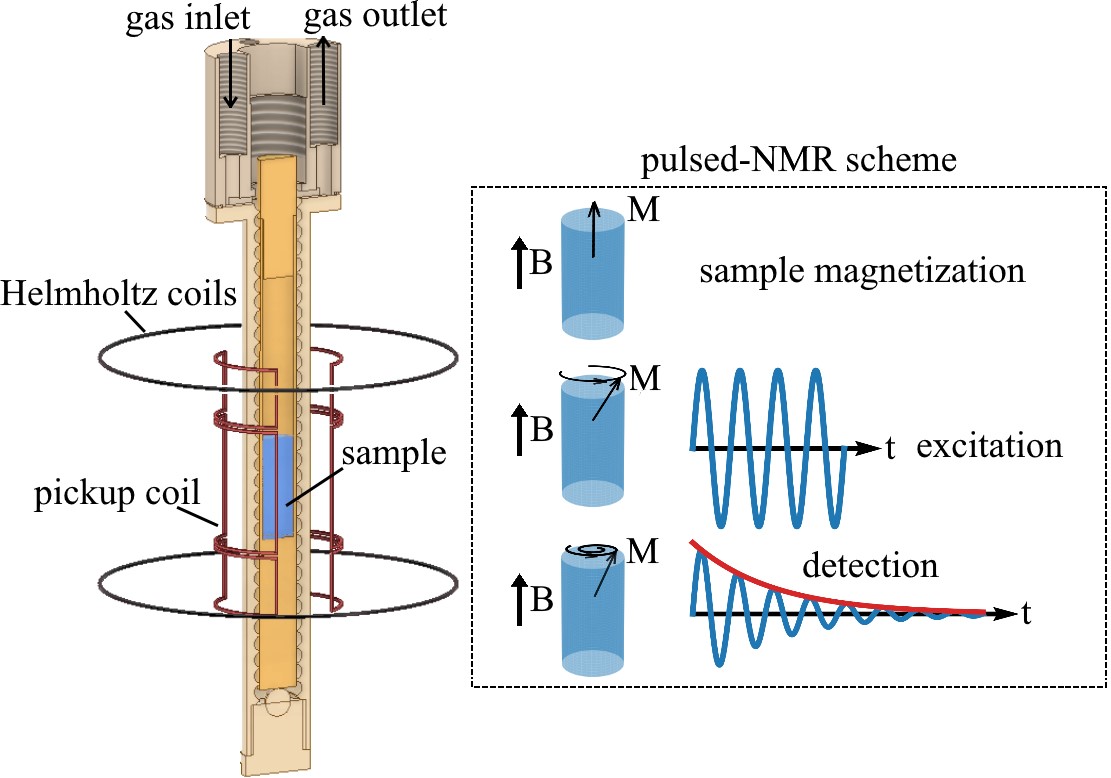}
    \caption{Left: the sample (blue) is at the center of the pickup coil (red).  
    The pickup coil consists of three Nb saddle-pair coils, where the area of the central coil is twice as large as the top and bottom ones. 
    Helmholtz coils (black) are used to fine-tune the bias field.
    Additional saddle-pair coils (not shown) are used to apply RF-pulses.
    The sample temperature is regulated and kept above the freezing point by flowing nitrogen gas through the spiral surrounding the sample holder.
    Right: pulsed-NMR scheme, used solely for characterizing the sample: the net magnetization vector $\mathbf{M}$ of the sample is initially along the bias field $\mathbf{B}$. 
    If an excitation pulse is sent, then $\mathbf{M}$ is tilted into the transverse plane and starts to precess. 
    The precessing $\mathbf{M}$ generates a signal detectable with the pickup and the SQUID.}
    \label{fig:setup}
\end{figure}

The sample of proton spins is thermally polarized by the bias field and acquire a net magnetization collinear to $\mathbf{B}_0$. 
To detect transverse magnetization of the sample, a niobium (Nb) pickup coil is mounted on the outside of the VTI in the liquid helium bath, see Fig.\,\ref{fig:setup}.  
The filling factor, defined as the volume ratio that the sample fills in the space enclosed by the central pickup coil, is approximately \SI{8}{\%}. 
A filling factor close to unity indicates that the dimensions of the sample match those of the pickup coil, optimizing the measurement sensitivity. 
However, due to the necessary vacuum insulation between the pickup and the sample, it is difficult to increase the filling factor. 
To reduce the noise, the cryostat features a built-in superconducting Nb shield.
A $\mu$-metal shield outside the cryogenic reservoir provides further magnetic shielding, and the entire apparatus is itself placed within a Faraday cage.

The magnetic flux generated in the pickup coil is coupled into a direct-current Superconducting Quantum Interference Device (dc-SQUID) operated in flux-locked-loop (FLL) mode.
The current in the SQUID input coil is converted into a feedback voltage, which is fed out of the cryostat and the Faraday cage via shielded copper Bayonet Neill-Concelman (BNC) cables into a Zurich Instrument MFLI Lock-in Amplifier (LIA) for processing and recording. 
With the excitation coils it is possible to send radio-frequency (RF) excitation pulses to conduct NMR experiments, while with the Helmholtz coils shown in Fig.\,\ref{fig:setup} we can generate an offset on the bias magnetic field to scan different frequencies. 

During measurements, the SQUID and FLL electronics were battery-powered to avoid electromagnetic interference (EMI). All other electronics are outside the Faraday cage, and only cables to power the excitation and Helmholtz coils are fed through the Faraday cage. One intrinsic noise source is due to the spins within the sample, known as spin-projection noise (SPN)\,\cite{Aybas_2021_Quantum_sensitivity}.
An overview of the parameters of CASPEr-Gradient-LF contributing to the sensitivity is listed in Table\,\ref{tab:parameters} for this measurement and future ones. 

\begin{table*}
\begin{threeparttable}
    \caption{\label{tab:parameters} 
    In addition to the measurement analyzed in this paper, examples of using \ch{^{129}Xe} and other samples in future measurements are considered. 
    The polarization here refers to molar polarization, the product of polarization and concentration of the target nucleus.  
    The polarization of methanol in this work is estimated from the proton thermal polarization at \SI{180}{\kelvin}. 
    There are numerous options for samples with polarization boosted orders of magnitude beyond thermal polarization. 
    }
    \renewcommand{\arraystretch}{1.3}  
    \setlength{\tabcolsep}{3pt}        
    \begin{ruledtabular}
    \begin{tabular}{p{0.23\linewidth}>{\centering\arraybackslash}p{0.23\linewidth}>{\centering\arraybackslash}p{0.23\linewidth}>{\centering\arraybackslash}p{0.23\linewidth}}
        \multicolumn{1}{c}{} & This work & \multicolumn{2}{p{0.46\linewidth}}{\centering Examples of future CASPEr-Gradient-LF} \\ \hline     
        Sample & Methanol & Methanol\tnote{a} &  Liquid \ch{^{129}Xe}\tnote{b} \\ \hline
        Spin density (\unit{\cm^{-3}}) & \num{5.9e22} & \num{5.9e22} & $\sim\num{1e22}$ \\ \hline
        Sample volume (\unit{\cm^3})& 1.2 & 1.2 & $\sim\num{1}$ \\ \hline
        Polarization technique & Thermal polarization & Brute Force\tnote{c} / PHIP\tnote{d} / DNP\tnote{e}  & SEOP\tnote{f} \\ \hline
        Polarization & \SI{1.8e-7}{} & $>\SI{1.1e-5}{}$ & $\sim\num{1}$\\ \hline
        $T_2$ (\unit{\s}) & 0.71 & 1 & 1000 \\ \hline
        $B_0$ inhomogeneity (\unit{\ppm}) & 10 & $\leq2$ & $\leq2$\\ \hline 
        \multirow{2}{=}{Maximum scan range} & \SI{1.348450}{\MHz} -- \SI{1.348690}{\MHz} & $\sim$\SI{1}{\kHz} -- \SI{4.3}{\MHz} & $\sim$\SI{1}{\kHz} -- \SI{1.2}{\MHz} \\ 
          & \SI{5.576741}{\neV/c^2} -- \SI{5.577733}{\neV/c^2} & $\sim$\SI{4.1}{\pico\eV/c^2} -- \SI{17}{\nano\eV/c^2} & $\sim$\SI{4.1}{\pico\eV/c^2} -- \SI{4.9}{\nano\eV/c^2} \\ \hline
        Electronic noise (\unit{\micro\Phi_0/\Hz^{1/2}}) & \SI{7.1}{} & \SI{1}{} &  \SI{1}{} \\ \hline
        Flux of SPN\tnote{g}\,\, (\unit{\micro\Phi_0}) &  $\num{.39}$ & $\num{.39}$   & \SI{0.16}{} \\ \hline
        \multirow{2}{=}{Spectral SPN in scan range (\unit{\micro\Phi_0/\Hz^{1/2}})} 
        & \multirow{2}{=}{\centering \num{0.11}} 
        & \SI{8.7}{} at \SI{1}{\kHz} 
        & \SI{3.3}{} at \SI{1}{\kHz} \\
        \rule{0pt}{0pt} 
        & \rule{0pt}{0pt} 
        & \SI{0.13}{} at \SI{4.3}{\MHz}  
        & \SI{0.11}{} at \SI{1.2}{\MHz} \\ \hline
        \multirow{2}{=}{$|g_{\mathrm{aNN}}|$ sensitivities in scan range (\unit{\GeV^{-1}})} 
        & \multirow{2}{=}{\centering \SI{3e-2}{}}  
        & \SI{1.0e-06}{} at \SI{1}{\kHz}  
        & \SI{1e-12}{} at \SI{1}{\kHz} \\
        \rule{0pt}{0pt} 
        & \rule{0pt}{0pt}  
        & \SI{1.2e-5}{} at \SI{4.3}{\MHz}  
        & \SI{1e-11}{} at \SI{1.2}{\MHz} \\
        
    \end{tabular}
    \end{ruledtabular}
    \begin{tablenotes}
    \footnotesize
    \item[a] Here we take \SI{180}{\kelvin} methanol prepolarized at \SI{2}{\tesla} and measurement time $100\,\tau_a$ for one scan step as the conditions for computing the baseline of future improvements. 
    \item[b] \ch{^{129}Xe}-enriched sample. The \ch{^{129}Xe} concentration can be close to unity. The details of calculations about liquid \ch{^{129}Xe} sample can be found in Ref.\,\cite{kimball2018overview}.
    \item[c] Prepolarization in a strong magnetic field and/or at low temperature. 
    \item[d] Parahydrogen-Induced Polarization (PHIP). 
    \item[e] Dynamic Nuclear Polarization (DNP).
    \item[f] Spin Exchange Optical Pumping (SEOP).
    \item[g] Spin projection noise (SPN). 
    \end{tablenotes}
    \end{threeparttable}
\end{table*}

\section{Measurement}

We searched for ALPs in a range of \SI{240}{\Hz} centered at \SI{1.348570}{\MHz} in steps of \SI{6}{\Hz}. The \SI{6}{\Hz} step size is approximately half of the NMR linewidth. 
At each step, the following actions were performed:
1) characterization of NMR sample and the setup, 
2) dwell time of \SI{5}{\s} for ring-down noise to decay,
3) data collection for \SI{100}{\s} to search for DM over $\approx$ 100 ALP coherence times,
4) ramping the bias field to change the Larmor frequency by \SI{6}{\Hz}, and 
5) another \SI{5}{\s} dwell time before repeating the steps. 
We performed three scans over the same frequency range to be able to cross-check potential signal candidates. 

To calibrate the sensitivity of the setup and to monitor the sample temperature, a pulsed-NMR measurement was carried out at each step. In such a measurement, the magnetization vector of the sample is tilted into the transverse plane by an RF pulse creating a magnetic field perpendicular to $\mathbf{B}_0$. 
The pulse frequency is chosen close to the Larmor frequency and the pulse duration is chosen such that a tipping angle of $\theta = \Omega_R\tau_p = \pi/2$ (usually called a $\pi/2$-pulse in NMR) is achieved. 
The Rabi frequency, $\Omega_R$, and the pulse duration $\tau_p=\SI{190}{\micro\s}$ were found by a pulse duration sweep.
The precession of the magnetization $\mathbf{M}$ after the pulse generates an oscillating signal in the SQUID (which is maximal for a $\pi/2$-pulse) that decays over time. 
The envelope of the decay signal is indicated by the red line in the pulsed-NMR scheme in Fig.\,\ref{fig:setup}. 
The transverse magnetization of the sample decays due to homogeneous and inhomogeneous relaxation mechanisms. 
The homogeneous relaxation time is called $T_2$. 
The total effective relaxation time due to all mechanisms $T_2^*\le T_2$ can be extracted from the linewidth of the spectral line: $T_2^*=1/(\pi\,\Delta\nu)$. Both $T_2$ and $T_2^*$ should be as long as possible to optimise the sensitivity in ALP searches\,\cite{Yuzhe2023_FrequencyScanning}. 

From the feedback voltage $V_\mathrm{f}$ of the SQUID, the magnetic flux at the SQUID input, $\Phi_{\mathrm{in}}$, as well as the magnetic flux inside the pickup coil, $\Phi_{\mathrm{pick}}$, can be reconstructed\,\cite{clarke2006squid}:
\begin{equation}\label{eq:squidpickup}
    V_\mathrm{f}
    = \frac{R_\mathrm{f}}{M_\mathrm{f}} \Phi_{\mathrm{in}}
    = \frac{R_\mathrm{f}}{M_\mathrm{f}} \frac{M_\mathrm{in}}{L_{\mathrm{pick}}+L_{\mathrm{in}}} \Phi_{\mathrm{pick}}
    = \varepsilon\,\Phi_{\mathrm{pick}}\,.
\end{equation}
Here, $R_\mathrm{f}$ is the feedback resistance and $M_\mathrm{f}$ is the feedback-coil mutual inductance. 
$L_{\mathrm{pick}}$ and $L_{\mathrm{in}}$ are the inductance of the pickup coil and SQUID input coil. 
The flux couples into the SQUID via a mutual inductance of $M_\mathrm{in}$.
The conversion factor is $\varepsilon \approx \SI{0.549}{\milli\volt/\Phi_0}$, where \unit{\Phi_0} denotes the magnetic flux quantum.

\begin{figure*}
    \centering
    \includegraphics[width=\linewidth]{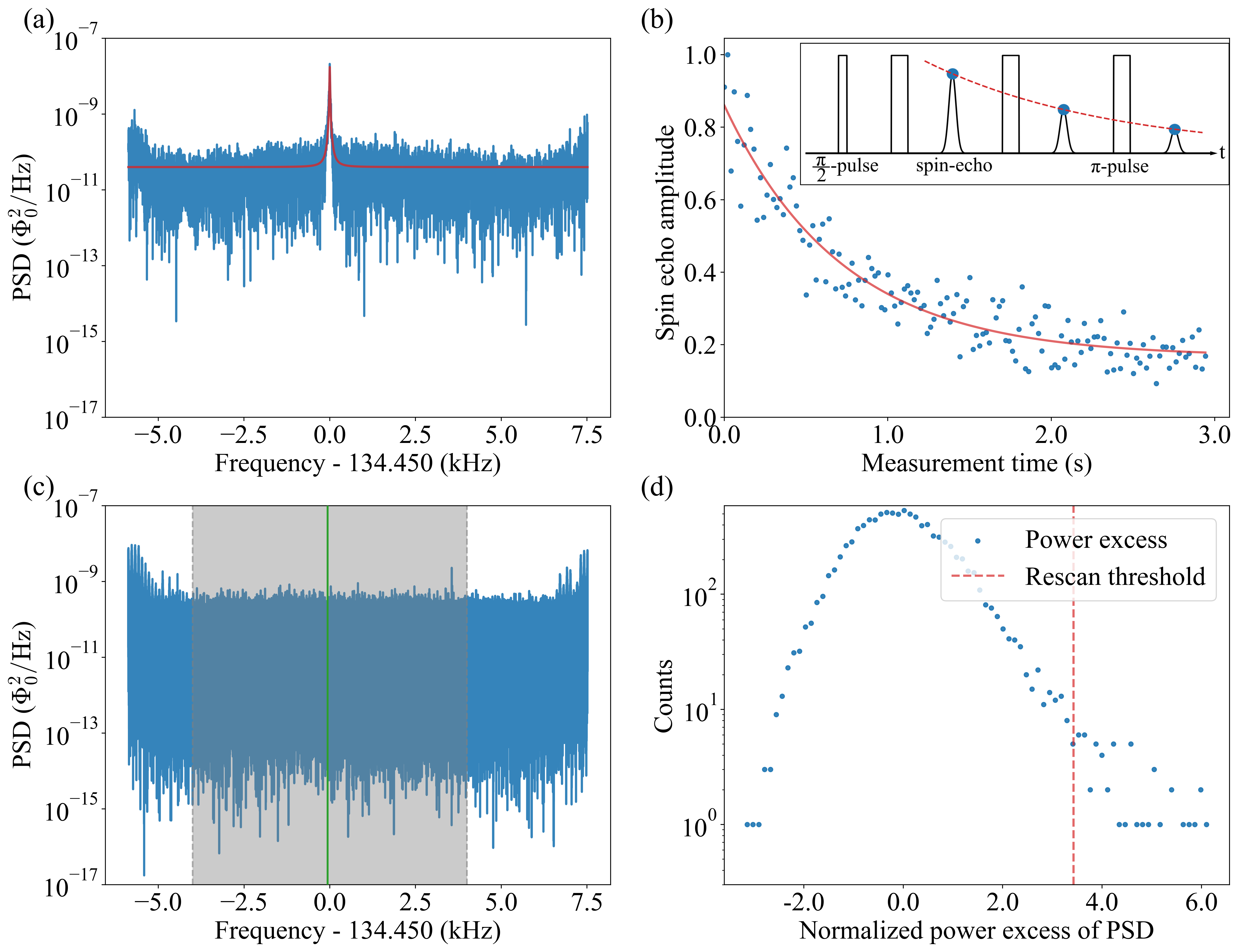}
    \caption{\textbf{(a)} PSD generated from the time-series of $\Phi_{\text{in}}$ after a $\pi/2$-pulse. 
    The NMR peak was fitted with a Lorentzian function (red line).
    The center of the peak is at \SI{1348449.21(5)}{\Hz} with a FWHM of \SI{14.79(1)}{\Hz}. 
    \textbf{(b)} CPMG-pulse-sequence measurement taken prior to the ALP-search steps. 
    From an exponential fit (red line) $y(t)=A e^{-t/T_2}+B$, where t is time and $y(t)$ is the normalized spin-echo amplitude (indicated by the blue dots), the relaxation time is extracted as $T_2 = \SI{0.72(5)}{\s}$. 
    \textbf{(c)} PSD of the first \SI{100}{s} ALP-search step.
    Structures at the lower and upper edges of the spectrum are artifacts from compensating for the built-in band pass filter of the LIA.
    The analysis window (grey) is centered at the Larmor frequency and has a width of \SI{8}{\kHz} to exclude the filter artifacts at the edges. 
    Parts of the spectrum outside of this window are not used in the data analysis. 
    We compute the noise baseline of each PSD within the analysis window. 
    For the spectrum shown, the average noise level is \SI{3.97(6)e-11}{\Phi_0^2/\Hz}. 
    We chose a resonant window (green) $\pm5$ NMR linewidths ($\pm \SI{70}{\Hz}$) around the  Larmor frequency to search for ALP-signal candidates around the Larmor frequency. 
   Outliers found outside the resonant windows can be used to examine outliers found on resonance in other data sets.
    \textbf{(d)} Histogram of the normalized power excess (NPE) generated from the PSD in (c) with the rescan threshold $\Theta_{\mathrm{RE}}$ (red line) at the \SI{95}{\%} confidence limit for a $5\,\sigma$-ALP detection. 
    }
    \label{fig:measurements}
\end{figure*}

In the LIA, the signal is demodulated into two channels, typically called the in-phase (I) and quadrature (Q) channels.
The output of the LIA, $I + i Q$ is used to generated the power spectral density (PSD) via discrete Fast Fourier Transformation (FFT).
An example of a PSD after applying a $\pi/2$-pulse is shown in Fig.\,\ref{fig:measurements}\,(a).
A Lorentzian fit to the PSD is used to extract the Larmor frequency, the full-width-at-half-maximum (FWHM) or linewidth $\Delta\nu$ and the power of the NMR signal.
These three parameters determine the sensitivity of the setup to the (pseudo)magnetic field (Eq.\,\ref{eq:pseudoB}).
Moreover, the sample temperature influences the linewidth. 
During the measurements, we did not observe significant changes in the linewidth; therefore, we conclude that the temperature was well regulated by the nitrogen gas flow.

We performed a Carr-Purcell-Meiboom-Gill (CPMG)\,\cite{CPMG} pulse sequence measurement to determine the $T_2$ relaxation time. A $\pi/2$-pulse followed by a series of $\pi$-pulses at fixed intervals (called echo time) is sent to the sample.
The resulting signal in the pickup coil features a series of spin echoes in between each two $\pi$-pulses (refocusing pulses). 
The spin-echo amplitudes decay exponentially with a decay constant $T_2$. 
We obtained $T_2=\SI{0.72\pm 0.05}{\s}$ for the sample from an exponential fit to the measured spin-echo amplitudes, see Fig.\,\ref{fig:measurements}\,(b). 
The average $T_2^*$ time for our setup obtained from NMR-peak fits is \SI{24.2 \pm 4}{\ms}. This places our setup in the regime where the ALP-coherence time $\tau_a=1/( \pi \Delta \nu_a )$, calculated from the linewidth of the ALP gradient broadening\,\cite{Gramolin2022Feb_SpectralSignatures} at \SI{1.348500}{\MHz}, is longer than $T_2^*$: $\tau_a \approx \SI{0.35}{\s}\gg T_2^*$.

After the characterization by pulsed-NMR and the CPMG pulse sequence, a dwell time of $\SI{5}{\s}$, which is longer than the characteristic longitudinal relaxation time $T_1$ ($\sim \SI{1}{\s}$), allowed time for the sample to return close to the equilibrium magnetization. 
Next, we recorded $V_\mathrm{f}$ for \SI{100}{\s} without sending pulses, with the lock-in frequency detuned by \SI{820}{\Hz} from the initial Larmor frequency\,\footnote{We detuned the lock-in frequency from the Larmor frequency by a few hundred hertz. 
Such a non-zero difference frequency presents a number of advantages.
For example, the demodulated signal oscillates at the difference frequency which, being a signature of the NMR signal, can be easily distinguished from noise.
}
The sampling frequency was \SI{13.4}{\kHz}. 
The data taken in this step were analyzed for possible signatures of ALPs. To prepare for the next scan step, we increased the Larmor frequency by \SI{\approx 6}{\Hz} (half of the NMR linewidth), either by ramping the superconducting magnet or the Helmholtz coils. 
The scan-step size was chosen such that the resonant windows would overlap. 
This overlap can allow the same ALP signal to appear in several neighboring scan steps, and can be used as a validity test\,\footnote{The strength of an ALP signal detected in multiple scan steps depends on the detunings of Larmor frequencies from the ALP Compton frequency. Such variation of ALP signal strength can be used to test the ALP signal candidates.}.
After changing the bias field there was another \SI{5}{\s} dwell time to ensure that the magnet power supply connection is physically interrupted in order to minimize ring-down noise in electronics.
We evaluated the PSD noise level over the measurement duration and found no effects caused by the ramping process.
The PSD of a full \SI{100}{\s} data set is plotted in Fig.\,\ref{fig:measurements}\,(c).
The resulting noise floor of our PSD spectra is determined by spin-projection noise, SQUID noise, electronic noise (e.g. the amplifier noise) and noise created by mechanical vibrations. 
The SPN in PSD is proportional to the number of spins and inversely proportional to the NMR linewidth [$\Delta\nu=1/(\pi T_2^*)$].
Since the NMR linewidth is proportional to the frequency, SPN in PSD scales inversely with frequency (see Table\,\ref{tab:parameters}). A typical noise spectrum of our setup is shown in\,Fig.\,\ref{fig:SQUIDfull} of the Appendix. 

Due to Earth's motion, the component of the ALP gradient perpendicular to $\mathbf{B_0}$ varies over time.
This leads to periodic modulations in signal strength, which affected the sensitivity over the duration of the measurements.
We consider the most sensitive scan as the main scan, and use the other two, which we will call A and B, as consistency checks.

\section{Data analysis}

We analyzed the data in a way similar to that of the HAYSTAC\,\cite{Brubaker2017_HAYSTAC}, the SHAFT\,\cite{SHAFT}, and the CASPEr-electric\,\cite{aybas2021_SolidStateNMR} experiments.
Each \SI{100}{\s} time-series corresponding to one individual scan step is converted to a PSD spectrum, as shown in Fig.\,\ref{fig:measurements}(c), which is then processed in three successive steps. 
First, features much narrower and wider than the expected standard halo ALP signal at the scanned frequency were smoothed out\,\footnote{This step renders the analysis insensitive to scenarios other than the standard halo model. 
For example, some models predict the existence of fine-grained DM streams in the galaxy \cite{Arza:2022dng, MiniclusterStreamsOHare, DALIstreams}. 
Such streams would feature a small velocity dispersion and would thus have spectral signatures narrower than what we consider in this work.}. 
Subsequently, a matched filter using the ALP lineshape expected for the gradient coupling is applied to generate normalized power excess.
Finally, after filtering the spectrum, bins with excess power greater than the threshold for a detection at $5\,\sigma$ significance are considered as outliers. 
Outliers near resonance are considered ALP signal candidates and undergo a scan-rescan test. 
In the following paragraphs we describe these analysis steps in detail. 

Savitzky-Golay (SG) filters\,\cite{SavitzkyGolay} are commonly used to smooth out features of a spectrum that are either narrower or wider than the expected signal. 
In the spectra, noise spikes can appear, for example, due to EMI penetrating the shields.
Additionally, impedance mismatch in the detection chain, composed of the pickup coil, the SQUID and the LIA, can lead to a broad and uneven baseline. 
We chose one half of the expected ALP linewidth at the scanned frequency as the filter window and applied a second-order SG filter to smooth out any features narrower than this.
If narrow features are present on top of an ALP signal, the signal lineshape can appear significantly distorted in the filter response, as we found by injecting simulated ALP signals into the data.
To minimize distortion, we refined the smoothing to selectively address only those narrow features that are unlikely to stem from white noise: 
the SG-filter response was subtracted from the unfiltered spectrum, leaving a residual spectrum containing only features narrower than the filter window.
The values in the (residual) PSD are expected to stem from a $\chi^2$ distribution with two degrees of freedom, if the time-series signals from the in-phase and quadrature channels of the LIA contain just Gaussian-distributed noise.
We performed $\chi^2$ fits to the histograms of the residual PSD, and determined the threshold for outliers. 
In the raw spectrum, if some PSD values differed from the filter response by an amount exceeding this threshold, they were identified as non-white-noise spikes.
Only these PSD bins as well as their four nearest neighbors were substituted with the average of the SG filter window, smoothing out those spikes that are likely EMI while preserving potential ALP signals.
We refer to the spectra after undergoing this spike suppression guided by SG-filter residuals as the spike-corrected spectra.
In all spike-corrected spectra of the main scan, \SI{0.002}{\%} of the total data points were modified in this way; in rescan A and B, \SI{0.004}{\%} and \SI{0.003}{\%} of points were modified.

Next, we performed a reduced chi-square test on the filtered spectrum to verify if the remaining noise is of purely random nature.
The test indicates the presence of a noise floor (baseline) that is not flat over the analysis window. 
To characterize the remaining noise, we fitted the spectrum baseline by applying another broad SG filter, this time with a filter window length of 200 ALP linewidths.
For normalization, this noise baseline is subtracted from the spike-corrected spectrum, then the result is divided by the baseline. 
An estimation of analysis efficiency considering all filtering steps is presented at the end of this section and is derived in more detail in the Appendix.

To search for ALP signal candidates in the spectra after baseline removal, we applied a matched filter (also known as optimal filter) using the expected ALP signal lineshape. 
For each scan step, we determined the spectral lineshape $\lambda(\nu,\,\nu_a)$ of the ALP signal as a function of frequency $\nu$ and ALP Compton frequency $\nu_a$.
The lineshape also depends on the continuously changing orientation of the sensitive axis of the setup with respect to the direction of Earth's motion through the DM halo\,\cite{Gramolin2022Feb_SpectralSignatures}.
Matched filtering is performed by convolving the PSD with the lineshape at a series of ALP Compton frequencies in the analysis window. 
The step size of the matched filter was chosen to be half the ALP linewidth. 
As verified empirically, such a step size not only maximizes the signal-to-noise ratio (SNR) but also reduces the analysis time compared to finer step sizes.
The result of matched filtering, normalized by its standard deviation, is referred to as normalized power excess (NPE).

To assess the likelihood of an ALP signal being present in the NPE, we conducted a hypothesis test using the NPE of each scan step as a test statistic.
Here we defined the presence of an ALP signal as the null hypothesis, and the absence of it as the alternative hypothesis.
We modeled the NPE at the ALP frequency bin for the two scenarios of a signal at $5\,\sigma$ significance being present, and not being present, by performing Monte-Carlo (MC) simulations (see Section\,\ref{sec:recoverytest} in the Appendix).
From the cumulative distribution functions (CDFs) obtained for both cases we located the NPE value $\Theta_{\mathrm{RE}}$ corresponding to the left-tail \SI{95}{\%} confidence interval limit of the null hypothesis, yielding a p-value of 0.001.
A histogram of the NPE is calculated with a bin width chosen according to the Freedman-Diaconis rule\,\cite{Freedman1981OnTH} to balance data resolution and noise, as shown in Fig.\,\ref{fig:measurements}\,(d).
In the NPE spectra, frequency bins at which the NPE value exceeds $\Theta_{\mathrm{RE}}$ were considered outliers.
\begin{table}[htb]
\centering
\begin{ruledtabular}
\begin{tabular}{lccc} 
\textbf{} & \textbf{Main scan} & \textbf{Scan A} & \textbf{Scan B} \\
\midrule
$N_\mathrm{full}$ & \SI{481680}{} & \SI{481680}{} & \SI{481680}{} \\
$ X_\mathrm{full}$ & \SI{6393}{} (\SI{1.32}{\%}) & \SI{5760}{} (\SI{1.20}{\%}) & \SI{6202}{} (\SI{1.29}{\%}) \\
$\langle X_\mathrm{full} \rangle$ & \SI{145}{} (\SI{0.03}{\%}) & \SI{145}{} (\SI{0.03}{\%}) & \SI{145}{} (\SI{0.03}{\%}) \\
\hline 
$N_\mathrm{res}$ & \SI{8430}{} & \SI{8430}{} & \SI{8430}{} \\
$X_\mathrm{res}$ & 5 & 7 & 3 \\
$\langle X_\mathrm{res}\rangle$ & 3 (\SI{0.03}{\%}) & 3 (\SI{0.03}{\%}) & 3 (\SI{0.03}{\%})\\
$X^{'}_\mathrm{res}$ & 5 & 5 & 3 \\
$X^{''}_\mathrm{res}$ & - & 0 & 0 \\
$\langle X^{''}_\mathrm{res}\rangle$ & - & 0 & 0 \\
\end{tabular}
\end{ruledtabular}
\caption{
Results of the three ALP scans. 
The expected values are derived assuming white noise. 
$N_\mathrm{full}$: number of bins in all NPE spectra.
$X_\mathrm{full}$: number of outliers (bins above rescan threshold). 
$\langle X_\mathrm{full} \rangle$: expected value of $X_\mathrm{full}$. 
$N_\mathrm{res}$: number of bins in resonant windows.
$X_\mathrm{res}$: number of ALP signal candidates (outliers within resonant windows). 
$\langle X_\mathrm{res} \rangle$: expected value of $X_\mathrm{res}$. 
$X^{'}_\mathrm{res}$: number of candidates after merging (candidates closer than one ALP linewidth are merged). 
$X^{''}_\mathrm{res}$: number of candidates appearing in this scan and also in the main scan within four ALP linewidths. 
$\langle X^{''}_\mathrm{res}\rangle$: expected value of $X^{''}_\mathrm{res}$. 
Values noted in parentheses following our results indicate the fraction of that quantity compared to the total bins, $N_{full}$.
}
\label{tab:candidates}
\end{table}

Table\,\ref{tab:candidates} lists the outliers and ALP signal candidates discovered over the investigated frequency range.
At every scan step the number of outliers, $X_\mathrm{full}$, exceeds the statistically expected number of false positives, $\langle X \rangle$, by at least an order of magnitude ($X \geq 10 \langle X \rangle$). 
The excess outliers in each spectrum were found to stem from non-white noise with linewidths neither narrow nor broad enough to be smoothed out by the SG filters. 
By distinguishing between resonant regions and backgrounds in the spectra, and by comparing the main scan to the rescans, we identified outliers resulting from noise.

Our setup is only sensitive to ALP interactions when the Larmor frequency of the sample spins is near the ALP Compton frequency. 
For that reason, out of the total number of outliers, $X_\mathrm{full}$, we considered only the $X_\mathrm{res}$ outliers within the resonant window to be candidates in need of further inspection.
In each scan step, candidates appearing less than one ALP linewidth apart (approximately \SI{1}{Hz}) were considered as a single candidate. This process reduced the number of candidates to $X^{'}_\mathrm{res}$.
The next step was to check whether any candidate in the main scan appeared in the other two scans as well. 
No such candidates were found; therefore, we conclude that no ALP gradient interaction was discovered at the detection sensitivity. 

In the absence of ALP-signal detection at the $5\sigma$ level, we calculate an upper limit for $g_{\mathrm{ap}}$:
\begin{equation}\label{eq:limcouple}
    |g_{\mathrm{ap,\, lim}} | = \sqrt{ \frac{ e_{5} } { \eta P_{\perp,\,1} } }\,.
\end{equation}
Here, $e_{5}$ denotes the signal power corresponding to the $5\sigma$ threshold of the NPE distribution. 
$P_{\perp,\,1}$ is a reference signal power expected for the case $|g_{\mathrm{ap}}|=\SI{1}{\GeV^{-1}}$. 
It is essential to correct the setup sensitivity and the coupling constants of ALP signal candidates for the efficiency of signal recovery, $\eta$.
We defined the efficiency of the analysis as the ratio of signal power recovered by the analysis to its original value, and we obtained with MC simulations $\eta=\SI{88.2\pm 4.6}{\%}$ (see Section\,\ref{sec:recoverytest}). 

\begin{figure}[ht]
    \centering
    \includegraphics[width=\linewidth]{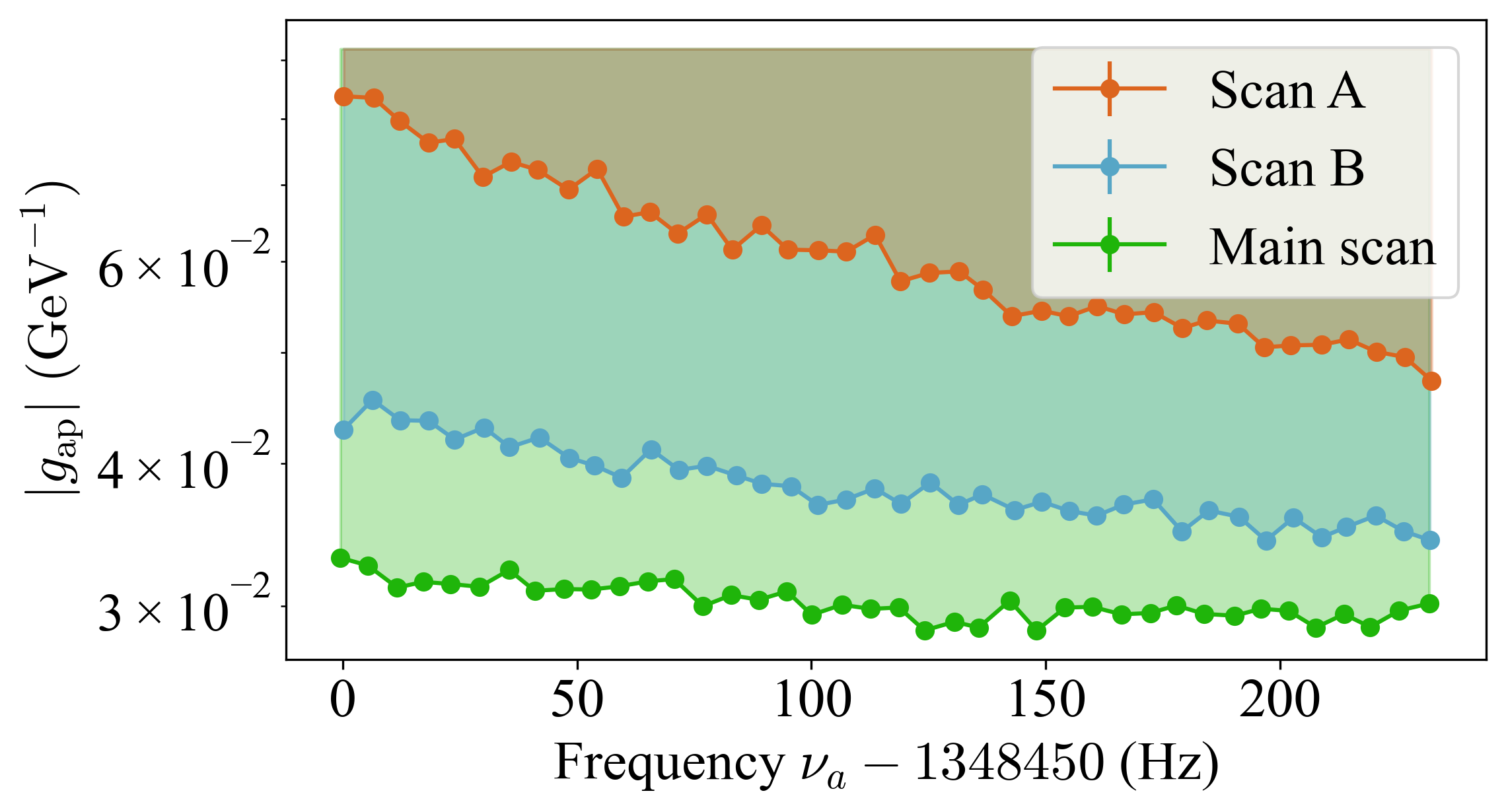}
    \caption{ALP gradient-proton couplings excluded at the \SI{95}{\percent} confidence level by the three sequential scans.
    Each data point represents the sensitivity to an ALP signal on resonance with the Larmor frequency of the respective scan step.
    Data taking happened within a time of approximately six hours, therefore the daily modulation of the ALP signal amplitude lead to an increase in sensitivity (see Fig.\,\ref{fig:modulation} in the Appendix).
    }
    \label{fig:ALPlimits}
\end{figure}

The limits obtained from our proof-of-concept measurements are plotted in Fig.\,\ref{fig:ALPlimits}.

\section{Conclusions and outlook}
We present an experimental search for galactic dark matter ALPs interacting with the protons of a thermally-polarized methanol sample employing SQUID-based detection of the transverse magnetization.
This is the first dedicated laboratory search in this general mass range. 
We scanned from \SI{5.576741}{\neV/c^2} to \SI{5.577733}{\neV/c^2} corresponding to a Compton frequency range of \SI{240}{\Hz} ($\approx$ 200 ALP linewidths) centered at \SI{1.348570}{\MHz}. 
We find no statistically significant evidence of ALP dark matter signals after performing a set of consistency checks.
This sets the ALP gradient-proton coupling $|g_{\mathrm{ap}}| \leq \SI{3e-2}{\GeV^{-1}}$ as the constraint in the scanned ALP mass range.
This narrow-bandwidth search serves as a commissioning test, and proof of principle for the methodology of CASPEr-Gradient-LF.

In ongoing experimental efforts, we employ various samples of higher nuclear-spin polarization. 
Those samples are polarized by techniques such as prepolarization in a strong magnetic field and low temperature (also referred to as ``Brute Force''), SEOP, PHIP and DNP\,\cite{Eills2023_Spin_Hyperpolarization}.
The polarization could reach the order of unity at maximum, boosting the sensitivity by more than six orders of magnitude compared to thermal polarization in this work. 

More methods are being developed to improve the sensitivity by up to four orders of magnitude, such as prolonging the relaxation time $T_2$ and increasing bias-field homogeneity.
Examples of using hyperpolarized samples with long relaxation times are given in Table\,\ref{tab:parameters}.
Increasing the volume of the sample can further improve the sensitivity.
The flexibility in the choice of sample is also advantageous to distinguish between various axion and ALP models that can have different couplings to protons and neutrons\,\cite{raffelt1999particle}.
With these upgrades, CASPEr-Gradient-LF is expected to probe the previously unexplored well-motivated ALP parameter space.




\begin{acknowledgments}

The authors acknowledge helpful discussions with Pablo Chrisvert, Jiawei Gao, Daniel Gavilan Martin, Grzegorz Łukasiewicz, Ophir Ruimi, Malavika Unni, and Pengyu Zhou. 
This work was supported in part by the Cluster of Excellence ``Precision Physics, Fundamental Interactions, and Structure of Matter'' (PRISMA+ EXC 2118/1) funded by the German Research Foundation (DFG) within the German Excellence Strategy (Project ID 39083149) and COST Action COSMIC WISPers CA21106, supported by COST (European Cooperation in Science and Technology), and also by the U.S. National Science Foundation under grant PHYS-2110388. 
The work of Y. Kim was supported by the Alexander von Humboldt Foundation. 
The work of A.O.S. and D.W.S. was supported by the U.S. National Science Foundation CAREER Grant No. PHY-2145162, and by the Gordon and Betty Moore Foundation, grant DOI 10.37807/gbmf12248. 
The work of O.M. was supported by the DFG, project FKZ: SFB 1552/1 465145163.

\end{acknowledgments}


\appendix


\section{Detection chain properties}\label{sec:squid}

\begin{figure}[ht]
    \centering
    \includegraphics[width=\linewidth]{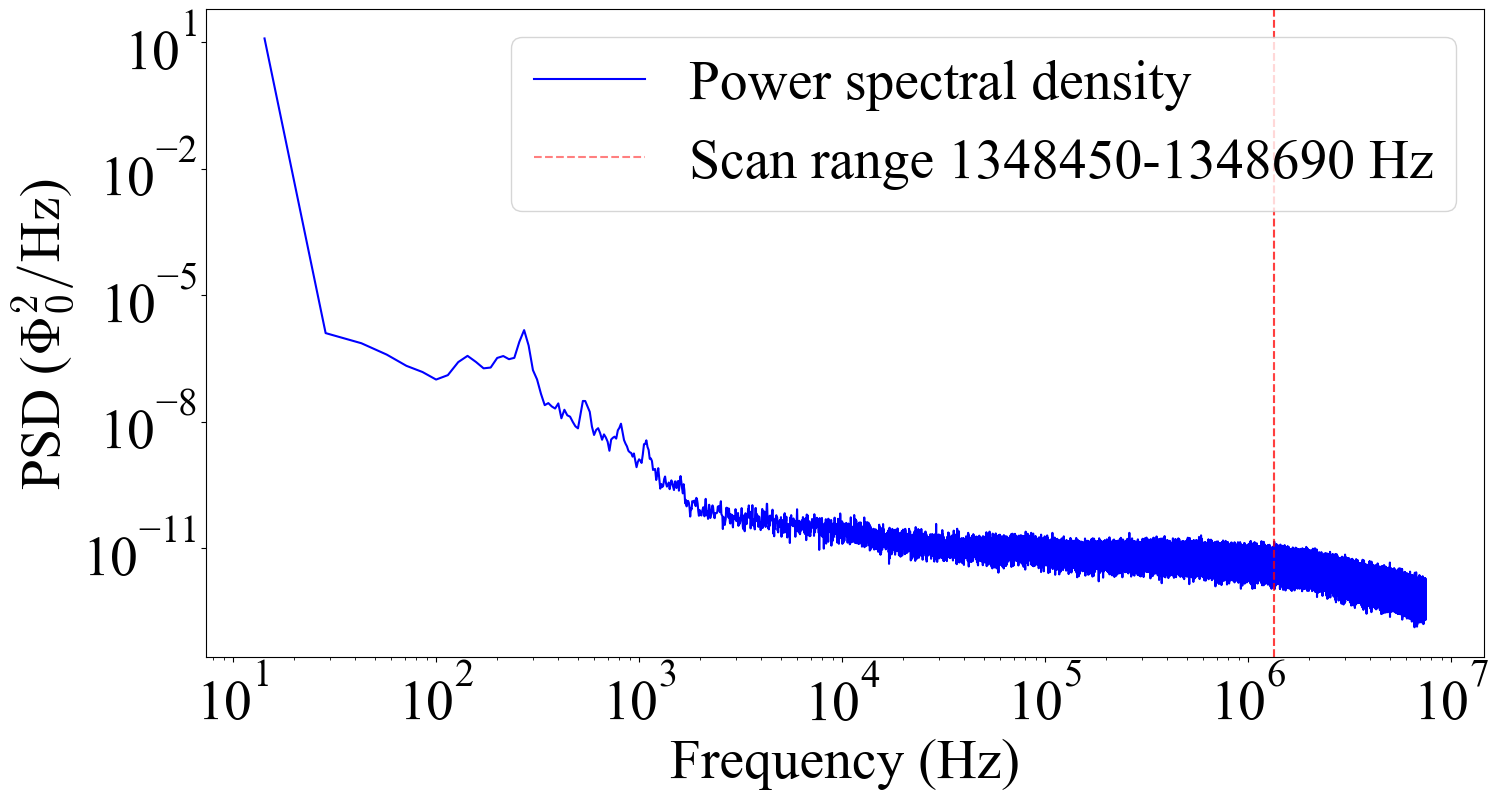}
    \caption{Spectrum showing the bandwidth of the SQUID, recorded with the bias field off and without sample. 
    The frequency probed in this work is marked by a dotted red line.
    The shown spectrum is an average of 10 spectra. }
    \label{fig:SQUIDfull}
\end{figure}

The properties of the dc-SQUID sensor used for this measurement are listed in Table\,\ref{tab:SQUIDtab}. 
The magnetic flux coupled into the SQUID, $\Phi_\mathrm{in}$, can be obtained from the feedback voltage signal, $V_\mathrm{f}$, via the relation:

\begin{equation}
    \Phi_{\mathrm{in}} = \frac{M_\mathrm{f}}{R_\mathrm{f}} V_\mathrm{f}\,,
    \label{eq:Phi_in_Vf}
\end{equation}
where $M_\mathrm{f}$ is the feedback-coil inductance and $R_\mathrm{f}$ is the feedback resistance. 
\begin{table}[h!]
\caption{Properties of SQUID sensor in use. Note that $R_f$ can be set to different values.}
\centering
\begin{tabular}{l  c }
\hline
\hline
Pickup coil inductance ($L_\mathrm{pick}$) & \SI{553}{\nH}\\[1ex]  \hline
SQUID input coil inductance ($L_\mathrm{in}$) &  400\,nH\,\\[1ex]  \hline
Mutual inductance ($M_\mathrm{in}$) &   \SI{03.98}{\nH}\,\\[1ex]  \hline
Feedback-coil inductance ($M_\mathrm{f}$) &   22.83\,\unit{\Phi_0}/mA\,\\ [1ex] \hline
Feedback resistance ($R_\mathrm{f}$) & \SI{3}{\kohm} \\[1ex]  \hline \hline
\end{tabular}
\label{tab:SQUIDtab}
\end{table}
Figure\,\ref{fig:SQUIDfull} shows the full PSD spectrum of the SQUID with the bias magnetic field turned off and without sample. 

We characterize the detection chain with the transfer coefficient $\beta$, which defines the relation between sample magnetization perpendicular to the bias field, $M_\perp$, and the amplitude of the oscillating feedback voltage of the SQUID, $V_\perp$: 
\begin{equation}
    \beta = \frac{V_\perp}{\mu_0 M_\perp}\,,
    \label{eq:beta}
\end{equation}
with the vacuum magnetic permeability, $\mu_0$. 
In the special case of $\pi/2$-pulsing, the sample magnetization vector is tilted into the transverse plane and $M_\perp = M_0 \sin(\pi/2)= M_0$ holds, where $M_0$ is the equilibrium magnetization of the sample.
For each scan step, we measured the feedback-voltage amplitude right after such a pulse, $V_0$, and calculated the equilibrium magnetization of the liquid methanol sample\,\cite{Abragam1961_NuclearMagnetism}:
\begin{equation}
    M_0 = \frac{n \gamma^2 \hbar^2 I(I+1) B_0}{3 k_B T}\,.
\end{equation}
Here, $\hbar$ is the reduced Planck constant, $k_B$ is the Boltzmann constant, $n$ is the number density of protons and I is the nuclear spin.
The proton is a spin $1/2$ particle with a gyromagnetic ratio of $\gamma=2\pi\times\SI{42.577}{\MHz/\tesla}$.
With four protons per methanol molecule, the sample contained \SI{7.2e22}{} protons within a cylindrical volume of \SI{1.2}{\cm^3}, with \SI{4}{\mm} radius and \SI{24}{\mm} height.
The sample at a temperature $T \approx \SI{180}{\kelvin}$ was placed within a bias field $B_0$ varying over the scans.
At a bias field of $B_0=\SI{0.0317}{\tesla}$ we find an equilibrium magnetization of $M_0\approx \SI{1.86e-10}{\tesla}/\mu_0$.
We computed the SQUID transfer coefficients for all scan steps and found an average value of $\beta\approx 4\times 10^6\,\mathrm{V/T}$.

\section{Expected axionlike-particle signal}\label{sec:ALPlineshape}
\begin{figure}[ht]
    \centering
    \includegraphics[width=\linewidth]{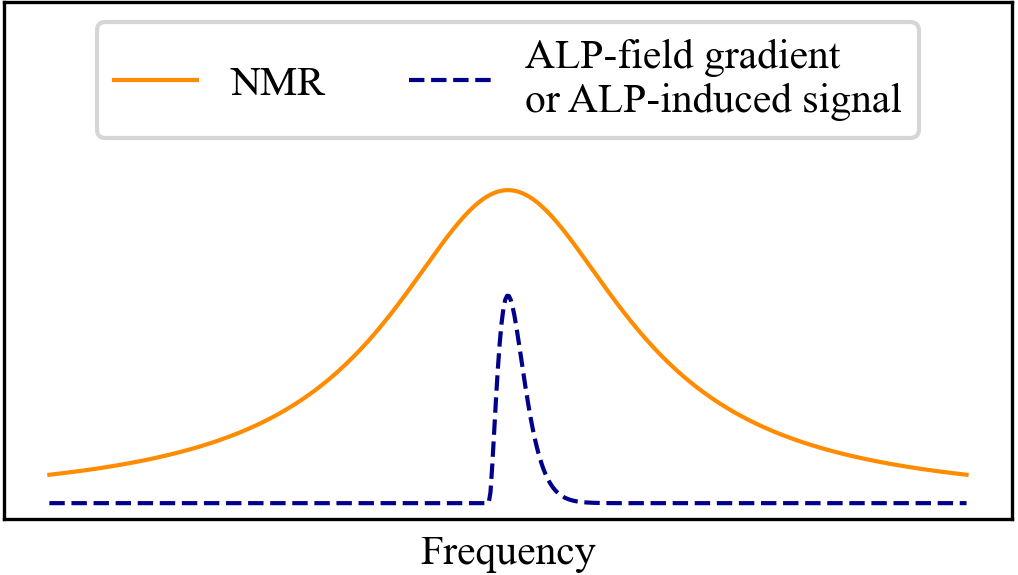}
    \caption{Illustration of spectral lineshapes of NMR, the ALP-field gradient and the ALP-induced signal. 
    In the shown scenario, where the NMR linewidth is much broader than the ALP-field gradient, only a small fraction of the spins are on-resonance, and the signal lineshape follows the ALP-field gradient.
    }
    \label{fig:NMR_ALP_lineshape}
\end{figure}

The magnetometer responds to the gradient of the ALP field, $\bm{\nabla} a$, which can be treated as a pseudo-magnetic field. 
The response has a linear dependence on $\bm{\nabla} a_\perp$, the component of $\bm{\nabla} a$ perpendicular to the bias field. 
In the measurements analyzed in this work, the NMR linewidth is ten times broader than the expected ALP linewidth, therefore the ALP-induced signal follows the spectral lineshape of $\bm{\nabla} a_\perp$ \cite{Yuzhe2023_FrequencyScanning}. The PSD of the signal is given by: 
\begin{equation}
    |S_{\perp}(\nu,\,\nu_a)|^2 = P_{\perp} \lambda_{\perp}(\nu,\,\nu_a)
    \,,\label{eq:Sperp}
\end{equation}
where $\nu$ is the frequency, $\nu_a$ is the ALP Compton frequency, $P_{\perp}$ is time-averaged signal power and $\lambda_{\perp}(\nu,\,\nu_a)$ is the normalized averaged spectral lineshape of $\bm{\nabla} a_\perp$ as found in Eq.\,(22) of Ref.\,\cite{Gramolin2022Feb_SpectralSignatures}. 
$\lambda_{\perp}(\nu,\,\nu_a)$ is illustrated in Fig.\,\ref{fig:NMR_ALP_lineshape}. 
In this work, we present PSDs in the units of \unit{\Phi_0^2/\Hz}, where \unit{\Phi_0} is the magnetic flux quantum.
$P_{\perp}$ can be derived as\,\cite{aybas2021_SolidStateNMR}:
\begin{align}
    \label{eq:axionpower}
    P_{\perp} &= \frac{1}{2} \left( \frac{M_\mathrm{f}}{R_\mathrm{f}}  \beta\,\mu_0 M_0\,  u \right)^2 \langle \xi^2\rangle \nonumber\\ 
    &=\frac{1}{2} \left( \Phi_{\pi/2} u \right)^2 \langle \xi^2\rangle\,.
\end{align}
Here, $\Phi_{\pi/2}$ refers to the magnetic flux in the SQUID after a $\pi/2$-pulse, converted from the feedback voltage by Eqs.\,\eqref{eq:Phi_in_Vf} and \eqref{eq:beta}. 
The spectral factor $u$, which is a function of Larmor frequency $\nu_L$, can be regarded as the effective fraction of on-resonance spins. 
$\xi$ is the tipping angle. 
In the following, we expand on the parameters in Eq.\,\eqref{eq:axionpower}. 

\begin{figure}[ht]
    \centering
    \includegraphics[width=\linewidth]{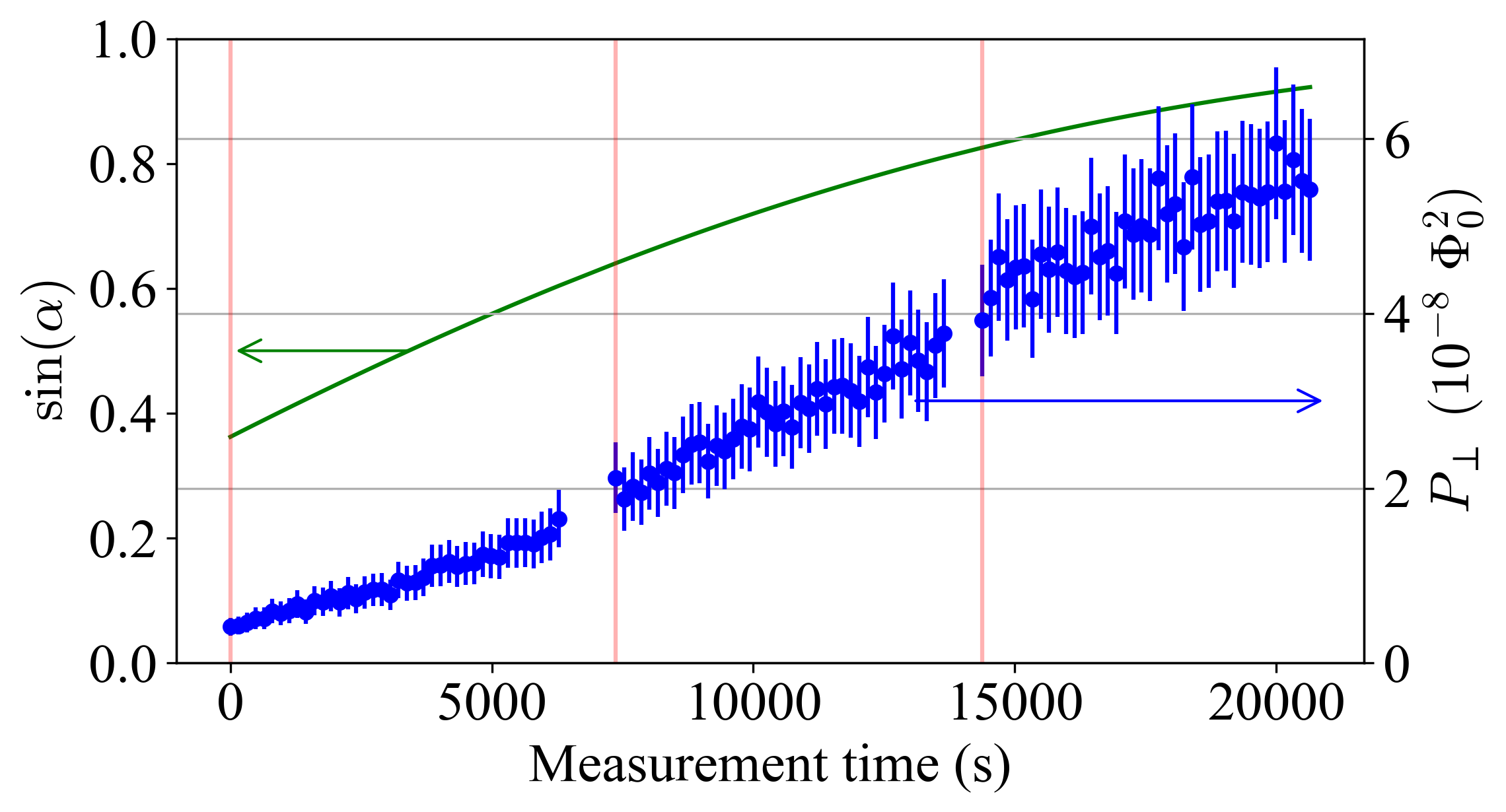}
    \caption{ALP parameters modulated by Earth's motion. 
    Plotted for the start times of each measurement step for each of the three scans in seconds. Zero time corresponds to start of the data taking on December 23, 2022, 15:47:30 (Central European Time, UTC+1). 
    Vertical dashed red lines mark the beginning time of a scan. 
    Green line: $\sin(\alpha)$ oscillates with a period of one day.
    Blue dot with error bar: expected ALP signal powers in the PSD, according to Eq.\,\eqref{eq:axionpower}, for the reference case of an ALP gradient coupling to proton of $g_\mathrm{ap}=\SI{1}{\GeV^{-1}}$.
    Uncertainties for these values were propagated from errors of the spin relaxation times $T_2$ and $T_2^*$, sample Larmor frequency $\nu_L$ and NMR signal power $P_\mathrm{NMR}$. 
    }
    \label{fig:modulation}
\end{figure}

The ALP-field gradient transverse to the bias field (acting as an oscillating pseudo-magnetic field) would slightly tip the sample magnetization from its equilibrium orientation.
The root-mean-squared (rms) tipping angle is determined by the rms Rabi frequency $\Omega_a$, relaxation time $T_2$ and the ALP-coherence time $\tau_a$\,\cite{Yuzhe2023_FrequencyScanning}:
\begin{equation}\label{eq:tipping_angle}
    \sqrt{\langle \xi^2\rangle} \approx \Omega_a T_2 \sqrt{\frac{\tau_a}{\tau_a + T_2}} = \left\{
    \begin{array}{cl}
    \Omega_a \sqrt{\tau_a T_2}\,,
    &  \tau_a \ll T_2 \\
    & \\
    \Omega_a T_2\,,
    &  T_2 \ll \tau_a
    \end{array} \right.
    \,.
\end{equation}
Note that as an implication of the stochastic nature of the ALP field, the Rabi frequency is not constant; therefore we use its rms value\,\cite{Centers2021Stochastic}:
\begin{equation}\label{eq:ALP_Rabi}
    \Omega_a \approx \frac{1}{2} \frac{c}{\hbar} g_\mathrm{aNN} a_0 m_a v_{\perp} 
    = \frac{1}{2} g_\mathrm{aNN} \sqrt{2\hbar c\rho_{DM}} v_{\perp}
    \,,
\end{equation}
with the speed of light $c$, the gradient coupling constant $g_\mathrm{aNN}$, the rms ALP-field amplitude $a_0$ and the ALP mass $m_a=\nu_a(2\pi \hbar/c^2)$.
$v_{\perp}= v_a \sin{\alpha}$ refers to the magnitude of the ALP velocity perpendicular to $\mathbf{B}_0$, where $\alpha$ is the angle between the ALP velocity, $\mathbf{v}_a$, and $\mathbf{B}_0$ (refer to Fig.\,1 of Ref.\,\cite{Gramolin2022Feb_SpectralSignatures}). 
$v_a$ is boosted by the velocity of the terrestrial detector relative to the galactic rest frame. 
This velocity is dominated by the galactic rotation speed at the solar radius, $v_{\odot}\approx\SI{233}{\km/\s}$. 
Earth's motion introduces annual and daily modulation into $v_{\perp}$. 
During the six-hour measurement, $v_{\perp}= v_a \sin{\alpha}$ varied by around \SI{50}{\%} due to the daily modulation of $\alpha$, which is plotted in Fig.\,\ref{fig:modulation}. 
Meanwhile, the magnitude of $v_a$ varied by less than \SI{1}{\%} around $v_{\odot}$ over the measurement duration. 
The factor of $1/2$ in Eq.\,\eqref{eq:ALP_Rabi} arises because only one of the two counter-rotating components of the linearly-polarized drive field is on resonance (rotating-wave approximation). 
In the spectral factor $u$, we consider the fraction of spins on-resonance with the ALP-field gradient. 
Considering the linewidth of NMR ($\Delta\nu_\mathrm{NMR}/\nu\approx\SI{10}{\ppm}$) and the ALP-field gradient ($\Delta\nu_a/\nu\approx\SI{1}{\ppm}$), we can roughly estimate the spectral factor to be $u\approx\SI{10}{\%}$. 
A more precise value can be computed by:
\begin{align}
    u=\frac{\Delta\nu_a}{\Delta\nu_\mathrm{NMR}} = \frac{T_2^*}{\tau_a}\,.
\end{align}
For the measurements around \SI{1.348500}{\MHz}, with $T_2^*=\SI{24.2 \pm 4}{\ms}$ and $\tau_a \approx \SI{0.35}{\s}$, the spectral factor $u$ is approximately $\SI{7}{\%}$. 

Summarizing the above contributions, we arrive at the expected rms ALP-signal amplitude:
\begin{align}
    |S_{\perp} (\nu,\,\nu_a)|^2_{\mathrm{rms}}& =\frac{1}{2} [\Phi_{\pi/2} u \Omega_{a} T_2]^2 \frac{\tau_a}{\tau_a+T_2} \lambda_{\perp} (\nu,\,\nu_a) 
    \,.\label{eq:S_perp_Tdelta}
\end{align}
We used synthetic ALP signals to test the recovery of signal candidate coupling constants, and to derive the sensitivity of the measurement, see section\,\ref{sec:recoverytest}.
Such signals were generated by evaluating the lineshape $\lambda_{\perp}(\nu,\,\nu_a)$ for a selected ALP Compton frequency $\nu_a$, scaling it with signal power $P_{\perp}(g_{ap})$ for a chosen coupling strength $g_{ap}$ according to Eqs.\,\eqref{eq:Sperp} and \eqref{eq:S_perp_Tdelta} and then adding (``injecting'') the generated signal on top of a PSD spectrum obtained from the measurements. 
If many spectra over the same frequency range are being averaged during data analysis, the expected ALP signal would be sufficiently described by the time-averaged lineshape $\lambda_{\perp}(\nu,\,\nu_a)$.
In our case, however, we look for signals in a single spectrum for each scan step.
The amplitude of the ALP field at any frequency is a stochastic variable with values drawn from a Rayleigh distribution.
The power that the ALP signal deposits in a single frequency bin of a PSD spectrum would be proportional to the square of its amplitude, and is therefore drawn from an exponential distribution\,\cite{Gramolin2022Feb_SpectralSignatures}.
To account for this stochasticity, we performed a bin-wise random scaling of $\lambda_{\perp}(\nu,\,\nu_a)$, where the PSD values of the ALP signal are multiplied by numbers drawn from an exponential distribution with unitary mean.
The ``smooth'' lineshape, $\lambda_{\perp}(\nu,\,\nu_a)$, would arise as the expectation value of the random-scaled ALP signal.

\section{Data analysis efficiency}
\begin{figure}[ht]
    \centering
    \includegraphics[width=\linewidth]{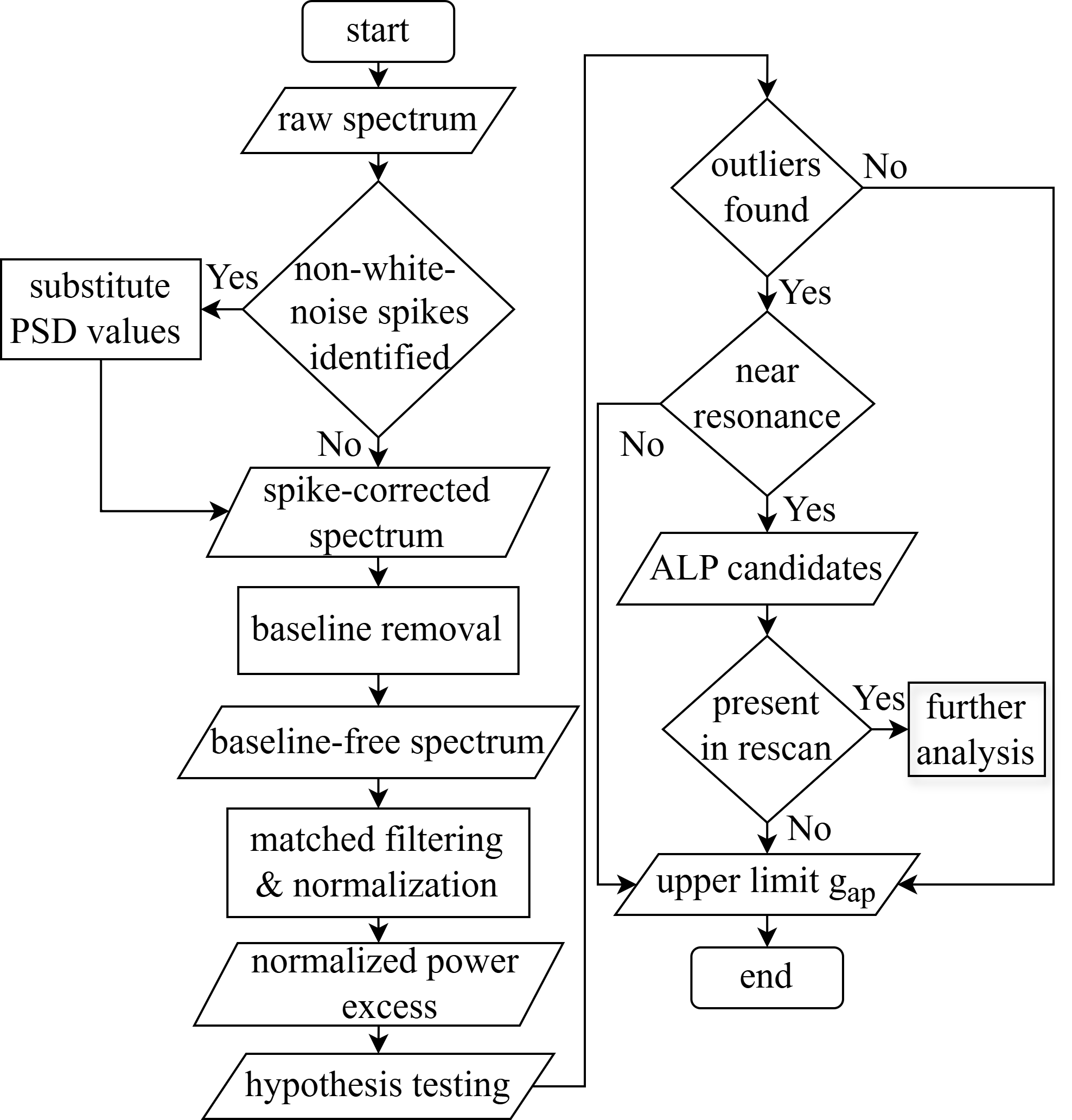}
    \caption{
    Schematic of the data-analysis procedure. 
    Non-white-noise-spike identification steps: Obtain the histogram of the residual spectrum (difference between raw spectrum and narrow SG-filter output). Calculate threshold from a $\chi^2_2$ fit to the histogram.
    Baseline removal: subtraction of and division by the broad-SG-filter output (baseline).
    Matched filtering: convolution of PSD with the ALP lineshape $\lambda_{\perp}(\nu,\,\nu_a)$. 
    Hypothesis testing: identification of outliers from the rescan threshold for an ALP signal at the $5\sigma$ level.
    ALP candidates: outliers within a window of $\pm5$ NMR linewidths around the Larmor frequency $\nu_L$.
    }
    \label{fig:flowchartDA}
\end{figure}

A schematic overview of the data analysis strategy reported in this work can be found in Fig.\,\ref{fig:flowchartDA}. 

\subsection{Data ``sanity'' check}\label{sec:recoverytest}
We injected artificial ALP signals after baseline removal. 
The recovery of signals injected after baseline removal serves as a useful check, demonstrating our ability to recover signals from an ideal spectrum with a flat baseline. 
In Section\,\ref{effectsofbase}, we examine how baseline removal affects the signal recovery and consider the overall efficiency of the data analysis.

We injected 20 different ALP signals with gradient couplings $\SI{0.001}{\GeV^{-1}} < g_\mathrm{aNN} < \SI{0.1}{\GeV^{-1}}$ into a raw PSD spectrum generated from data taken in one of the scan steps.
All artificial ALP signals were introduced at Compton frequencies randomly chosen from the analysis window.
Injections and subsequent data analyses were carried out individually so that each spectrum used to test signal recovery only contained a single synthetic ALP signal.
\begin{figure}[!h]
    \centering
    \includegraphics[width=\linewidth]{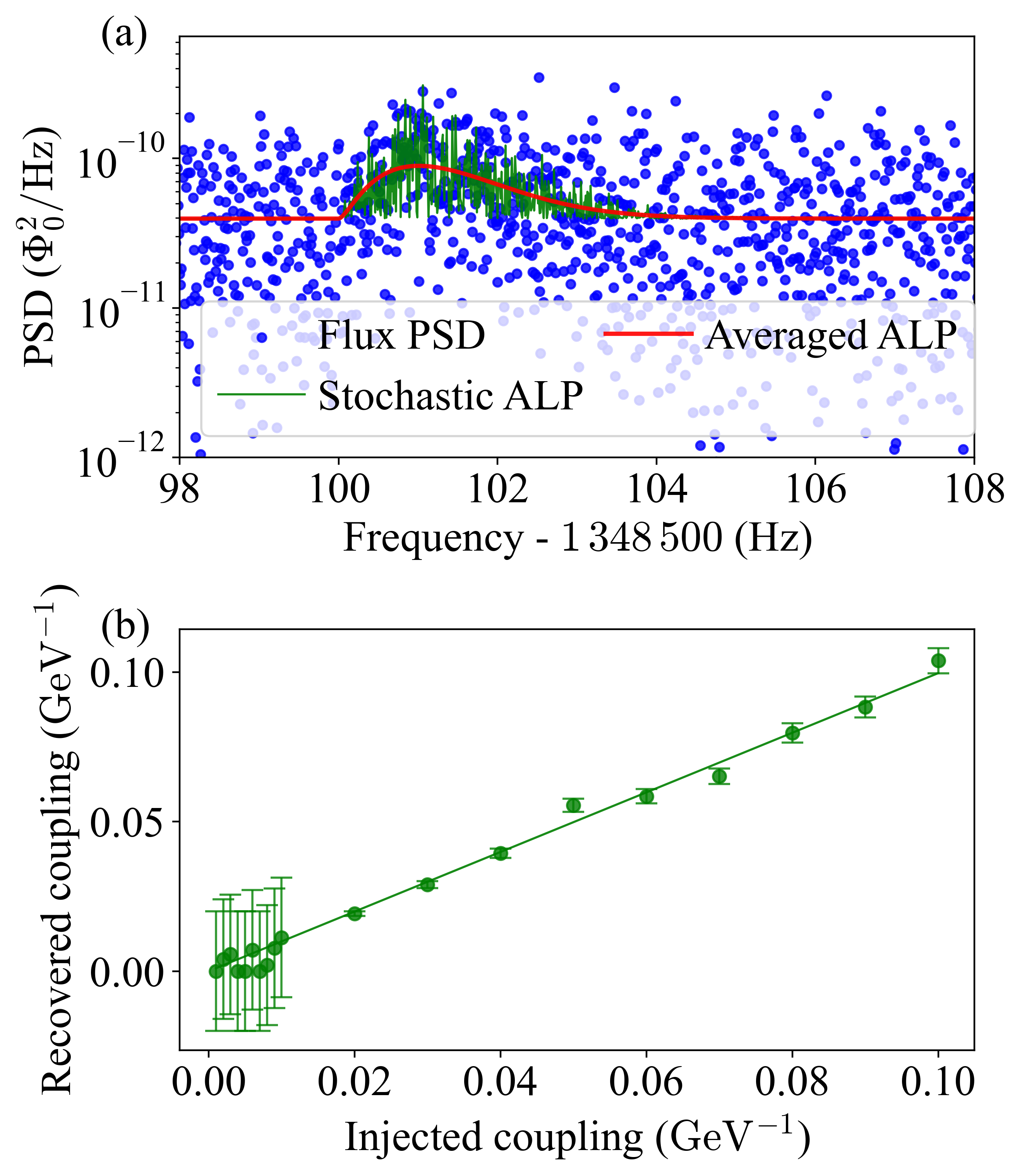}
    \caption{
    (a) Zoomed-in view of the spectrum of one of the scan steps of the main scan. 
    The PSD values are plotted in blue. 
    Shown in red is the time-averaged signal lineshape $\lambda_{\perp}(\nu,\,\nu_a)$ of one of the injected ALP signals, shown in green is $\lambda_{\perp}(\nu,\,\nu_a)$ after random scaling with values drawn from an exponential distribution.
    (b) Gradient coupling constants recovered from 20 ALP signals, injected at random frequencies after baseline removal had been performed. 
    Plotted 
    are the signal power values of the matched filtered PSD at the ALP Compton frequencies converted into gradient coupling. 
    Note that for couplings below the setup sensitivity, we indicate the uncertainty of the recovered value with the $5\sigma$ sensitivity interval.
    }
    \label{fig:fakeALPsensitivity}
\end{figure}Figure\,\ref{fig:fakeALPsensitivity}\,(a) showcases one synthetic ALP signal injected at a Compton frequency of $\nu_a=\SI{1348600}{\Hz}$, with a gradient coupling of $g_\mathrm{ap}=5\times10^{-2}\,\mathrm{GeV}^{-1}$.
This injected ALP signal is recovered by data analysis with a coupling of $\SI{4.8e-2}{\GeV^{-1}}$ with an uncertainty of a few percent.
The recovered coupling constants of all injected ALP signals are plotted in Fig.\,\ref{fig:fakeALPsensitivity}\,(b).
For the flat baseline case, an agreement between the injected and recovered coupling is expected, and in fact, we found a linear relation with a slope of \SI{0.999\pm0.007}{}.

\subsection{Effects of baseline removal}
\label{effectsofbase}

The effect of baseline removal depends on the shape of the baseline and, in our case, on the chosen window length of the SG filter.
The SG filter used to obtain the baseline may accidentally fit potential ALP signal peaks as noise.
When the filter window is sufficiently large, this can produce a significant shift in the NPE distribution obtained after filtering, compared to a scenario without baseline removal. 
This shift in NPE leads to a degradation of accuracy in recovering the signal coupling strength from NPE, which needs to be taken into account when evaluating the efficiency of the data analysis.
We define the signal recovery efficiency including baseline removal, $\eta_{\mathrm{rec}}$, as follows:
\begin{equation}
    \eta_{\mathrm{rec}} = \frac{e_{\mathrm{rec}} - \langle e_{\mathrm{off}} \rangle}{e_{0}}\,.
\end{equation}
Here, $e_{\mathrm{rec}}$, $e_{0}$, and $\langle e_{\mathrm{off}} \rangle$ are the average NPE values at the frequency of an ALP signal with SG filter, without SG filter, and the offset due to the SG filter, respectively.

To compute the average efficiency, we used the baselines obtained via broad-SG filtering from ten of the PSD spectra.
For each of these baselines, we generated 1000 sets of random noise data following a $\chi^2$ distribution with two degrees of freedom and with the mean value adjusted to zero. 
These data sets simulate baseline-free, normalized spectra containing purely white noise (and are referred to as pure-noise spectra).
An artificial ALP signal was injected into all data sets, with a coupling strength adjusted to ensure a 5-$\sigma$ significance in the absence of a baseline.
Then, PSDs with non-flat baselines were simulated by combining the pure-noise data sets with the baselines from the measurements, as \textit{baseline $\times$ (1 + noise)}.
Both sets of PSDs, with and without baseline, underwent the analysis procedure to obtain their NPE.
Finally, we constructed the histogram of NPE values recovered in the bin containing the frequency of the injected ALP across all simulated PSDs.
Similarly, the distribution of NPE values in the absence of an ALP signal was collected from a frequency bin more than five ALP linewidths remote from the ALP frequency.
\begin{figure}[ht]
    \centering
    \includegraphics[width=\linewidth]{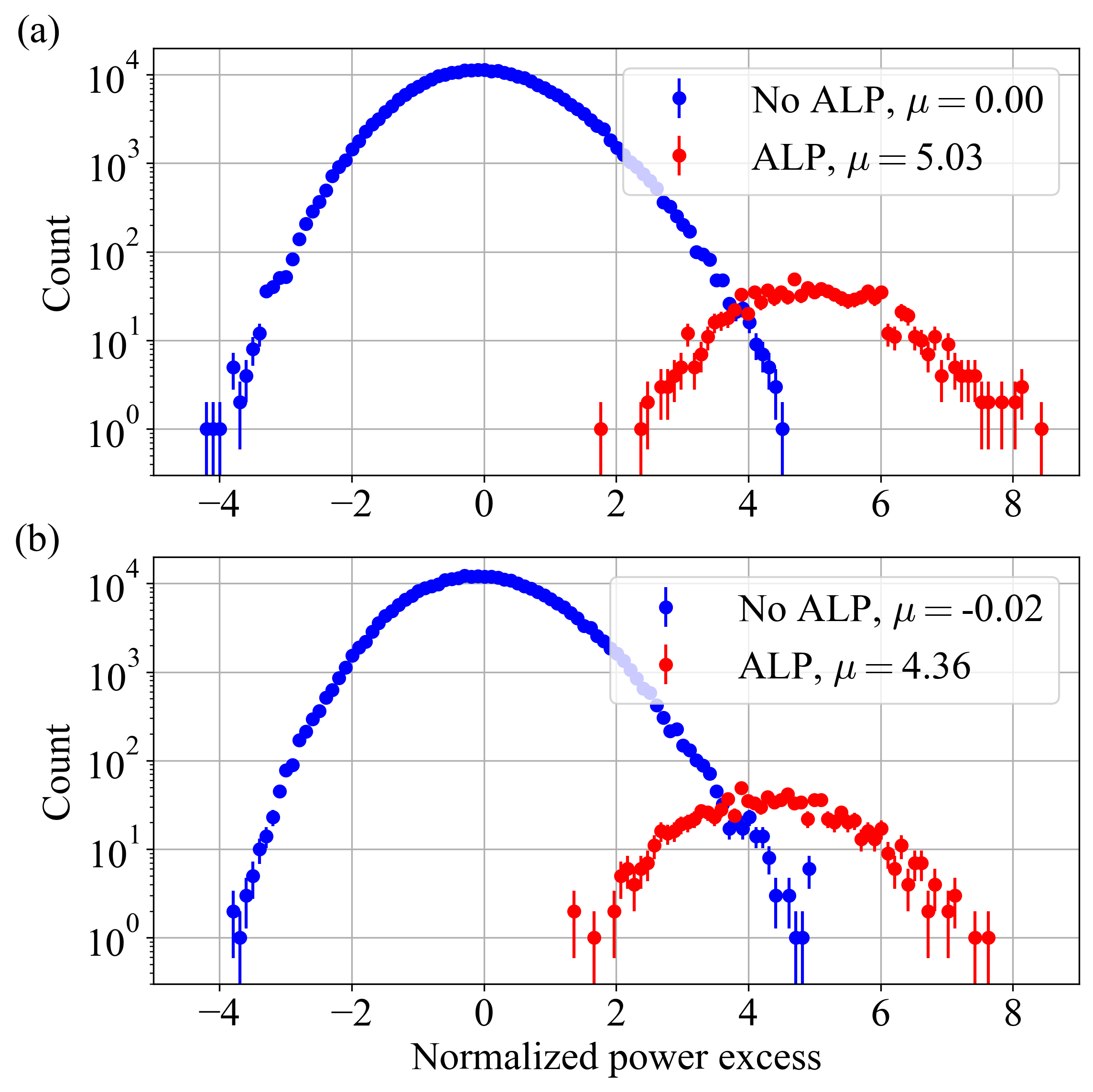}
    \caption{Histogram of normalized power excess (pull distribution) with ALP (red) and without ALP (blue). 
    \textbf{(a)} Flat baseline. 
    \textbf{(b)} With real baseline and SG filter. 
    Here, $\mu$ refers to the mean of a histogram. 
    }
    \label{fig:pulldistr}
\end{figure}
Figure\,\ref{fig:pulldistr} depicts the NPE histograms for these scenarios: flat baseline, obtained from the pure-noise spectra, and non-flat baseline, obtained from the noise spectra combined with real baselines; with and without ALP signal for both cases.
For a flat baseline [Fig.\,\ref{fig:pulldistr}\,(a)], the NPE values recovered from the ALP signal bin (red) are distributed around a mean of $e_0=5.0$, consistent with the signal-to-noise ratio of the injected signal. 
In contrast, for spectra where the SG filter has removed a non-flat baseline [Fig.\,\ref{fig:pulldistr}\textbf{(b)}], the distribution of ALP signal NPEs is shifted, with a mean of $e_{\mathrm{rec}}=4.36$.
The average offset due to the effect of the filter, $\langle e_{\mathrm{off}} \rangle$, is obtained from the difference between the no-ALP NPE histograms with and without baseline.
We arrive at an average recovery efficiency $\eta_{\mathrm{rec}} =\SI{88.2\pm 4.6}{\%}$. 
In terms of signal coupling strength, the efficiency is $\sqrt{\eta_{\mathrm{rec}}} =\SI{93.9\pm 2.5}{\%}$ and has been taken into account in the calculation of the sensitivity limits.

\subsection{Finding the rescan threshold}

Previously, we obtained the expected distribution of NPE values depending on the presence of an ALP signal, through Monte-Carlo (MC) simulations. 
To determine whether the NPE of each frequency bin of a spectrum obtained from the experiment is likely to be a signal from ALPs or a statistical fluctuation, we performed a hypothesis test.

\begin{figure}[ht]
    \centering
    \includegraphics[width=\linewidth]{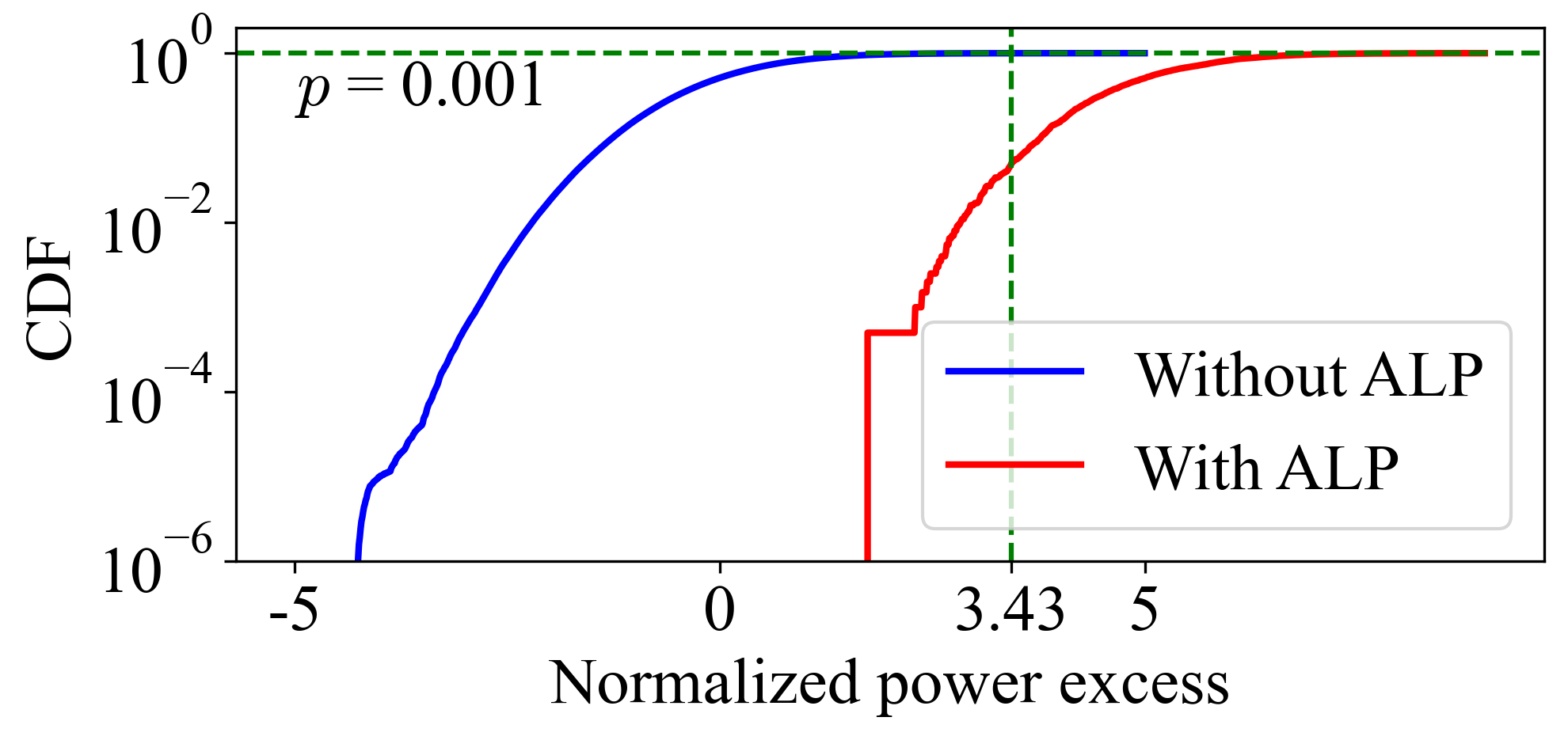}
    \caption{The empirical CDF obtained from the NPE distribution of spectra with a flat baseline [refer to Fig.\,\ref{fig:pulldistr}(a)].
    The vertical dashed line represents the threshold for rescan, $\Theta_{\mathrm{RE}}$, corresponding to the null hypothesis of a $5\,\sigma$ ALP detection at the \SI{95}{\%} confidence level.
    The p-value corresponding to the threshold (horizontal dashed line) is approximately 0.001. 
    }
    \label{fig:Hypothesis test}
\end{figure}

In the simulation, we assumed an ALP with a coupling strength corresponding to $5\,\sigma$.
This signal was used as the null hypothesis. 
From it, we found the NPE providing a \SI{95}{\%} confidence level. 
We calculated the empirical cumulative distribution function (CDF) using the histogram of NPE values at the ALP signal frequency bin obtained through MC simulations for the case of a flat baseline, i.e., the ideal case without using an SG filter.
Figure\,\ref{fig:Hypothesis test} shows the empirical CDF obtained from MC simulation.
The threshold value at the \SI{95}{\%} confidence interval limit was found to be $\Theta_{\mathrm{RE}}=3.43$. 
In the experiment, frequency components that yielded an NPE greater than this were considered signal candidates or outliers, depending on whether they were on resonance or not.
ALP signal candidates from the main scan were compared with scans A and B in order to determine if they were signals from statistical fluctuations.


\bibliographystyle{unsrt}
\bibliography{references}

\end{document}